# Phenomenology for Supernova Ia Data Based on A New Cosmic Time

**Charles B. Leffert**   Wayne State University, Detroit, MI 48202 USA
Email: C_Leffert@wayne.edu

**Abstract** A new phenomenological theory for the expansion of our universe is presented. Because fundamental supporting theory is still in development, its discussion is not presented in this paper. The theory is based on a new algebraic expression for cosmic time $G\rho t^2 = 3/32\pi$, which correctly predicts the WMAP measured cosmological constants and the fundamental Hubble parameter $H(t)$ for the expansion of the universe. A replacement for dark matter, called here "dark mass", is proposed which scales as $R^{-2}$ with the expansion and incorporated. It does not react with ordinary matter, except gravitationally, and produces flat rotational curves for spiral galaxies. Also a new expression for the approaching velocity of radiation in a closed 3-sphere expanding universe is given that accounts for the early degrading negative approach of radiation for z > 1.7. The expression is $v_c = Hr - c$. Combining these three elements produces a luminosity distance $d_L$ that successfully predicts the apparent magnitude of exploding supernova Ia stars and even the new gamma ray bursts with no need for dark energy or acceleration of the expansion of the universe.

> Nevertheless, this is the only level of explanation that classical cosmology offers: the universe expands now because it did so in the past. Although it is not included in the list, one might thus with justice add an 'expansion problem' as perhaps the most fundamental in the catalogue of classical cosmological problems. Certainly, early generations of cosmologists were convinced that some specific mechanism was required in order to explain how the universe was set in motion.
> From: *Cosmological Physics* BY John A. Peacock
> @1999 by Cambridge University Press. Used with permission.

## 1.0 Introduction

Measurements of the flux of radiation from the luminosity (L) of supernova Ia class of stars (SNIa) that exploded in the distant past offer astronomers a tool for exploring the precise history of how our universe has been expanding in time in terms of the fundamental scale factor $R(t)$, or Hubble parameter $H(t)$, where t is the age of the universe. But a correct theoretical $H(t)$ depends on the contents of our universe as well as how the density of each constituent scales with the expansion of our universe.

The main current theory is that of general relativity (GR) using the Friedmann equations with an added period of inflation and an added constituent called *dark matter*. The early 1998 higher redshift (z >1) SNIa measurements [1, 2] did not quite fit the theoretical predictions. Theorists then added another constituent *Einstein's lambda* or *dark energy* to our universe to improve the fit of theory to the data.

The postulated *dark matter* has never been identified, but it is described as having a scaling factor $R^{-3}$ with the expansion of our universe. The only understanding of the *dark energy* is that somehow it causes the acceleration of the expansion of our universe. Even with the addition of the dark energy, the scatter of the measurements of SNIa data has been too large to determine the scaling of this dark energy with the expansion. So the $H(t)$ is not yet fixed and GR-theory is questioned. The inadequacy of the present Friedmann equations will be shown in Sections 3 - 6.

Under the premise that general relativity and the Friedmann equations are not correct **globally** and with this author's new concepts of the beginning of our universe and the nature of its energy and gravity to explore, cosmological calculations were begun (1990) [3, 4]. The very first model showed promise and soon led to an algebraic expression that fixed the expansion Hubble parameter $H(t)$ and the current cosmological constants $H_0$; densities of radiation, $\rho_{r0}$ and mass, $\rho_{m0}$; and present age of the universe, $t_0$. All of them had values within the measured range of uncertainty.

As new concepts multiplied in the search for a solid theoretical foundation of the beginning and source of the expansion, papers were written [5, 6] and referred to in the arXiv papers as "Spatial Condensation or SC-theory", which was quite different from current physics. On the other hand, the algebraic expression for the expansion, after the beginning of our universe, is so fundamentally simple and complete, that it was clear that it should be made available to the scientific community in a refereed journal. Therefore without the theoretical foundation, it is presented here as a phenomenological representation for the expansion of our universe

In Section 2 of this paper, the integral, universal expansion function Eq. (1) is deduced from first principles and basic physical assumptions; but for presentation as phenomenology, the description of the new even more basic foundational concepts that led to those assumptions are omitted. In Section 3 the working relations of the cosmological parameters, from differentiation of Eq. (1) and listed in the Appendix, are shown for the expansion theory that is complete and has no free parameters. Before discussing the application to SNIa theory, Section 4 presents predictions of the cosmological constants and new "dark mass" rotational curves of spiral galaxies. Section 5 develops theory to include the transport of radiation in an expanding universe, Section 6 extends the theory specifically to Supernova Ia. Section 7 includes gamma ray bursts and section 8 considers other features of the new cosmic time. The Summary and Conclusions are presented in Section 9.
.
**2.0 Derivation of the New Universal Expansion Time**

Our universe has contents of radiation, baryonic matter and dark **mass** (not matter). In SC-theory, as in GR-theory, radiation scales as $R^{-4}$ and matter scales as $R^{-3}$. However, an important first SC-assumption is that dark mass scales as $R^{-2}$. An important second assumption is that SC-geometry is a closed 3-sphere, as assumed by Einstein in his first attempt at GR-cosmology [7], except that whereas Einstein assumed a static universe with only three-spatial dimensions, an expanding universe is assumed here of radius $R(t)$ embedded in a higher dimensional space. The density of this new SC-dark mass decreases with the expansion but its total mass in the universe continues to grow with the expansion.

The algebraic expression for the expansion is the complete relation between cosmic time t and both density $\rho$ and radius $R$,

$$G\rho t^2 = \kappa = 3/32\pi . \tag{1}$$

With present values having subscript 0, the SC total density postulated here is,

$$\rho(R) = \rho_{r0}(R_0/R)^4 + \rho_{m0}(R_0/R)^3 + \rho_{x0}(R_0/R)^2 . \tag{2}$$

For later use, the derivative with respect to $R$ of $\rho$ is,

$$\rho' = \partial\rho/\partial R = -2\rho_2/R , \tag{3a}$$

and $\dot{\rho} = -2H\rho_2$ so that,

$$\rho_2 = 2\rho_r + (3/2)\rho_m + \rho_x . \tag{3b}$$

The GR Friedmann equation with $k = \Lambda = 0$, p neglected and $\rho$ the total GR density (using the first two terms of Eq. (2) with $\rho_{m0}$ including dark matter) is,

$$H^2 = (8\pi G/3)\rho . \tag{4}$$

This GR expression for $H^2$ is proportional to $\rho$, and $\rho \to 0$ drives $H \to 0$ as it should. But in the new theory, an SC-equation is needed with a dimensionless factor that greatly strengthens the dependence on the slowly changing contents of domination from radiation to matter, to dark mass with its new $1/R^2$ dependence, and finally to an effectively empty universe. The expression for the dimensionless factor needed can be obtained from Eqs. (2) and (3).

Note that $\rho/\rho_2$ from Eqs. (2) and (3) form such a factor and in the limit $R \to \infty$, $\rho/\rho_2 \to 1$. Including this factor squared modifies Eq. (4) to:

$$H^2 = (G/\kappa)(\rho/\rho_2)^2 \rho , \tag{5}$$

where $\kappa$ is a new constant and $\rho$ now includes the $1/R^2$ term, $\rho_x$ given in Eq. (2). Taking the positive root of both sides gives,

$$H = \dot{R}/R = (G/\kappa)^{1/2}(\rho/\rho_2)\rho^{1/2} . \tag{6}$$

One can verify that $(G\rho/\kappa)^{1/2} = 1/t$ is a solution (see also Section 2.1) that on substitution in Eq. (6) gives,

$$tH = \rho/\rho_2, \tag{7}$$

Squaring the solution gives,

$$G\rho t^2 = \kappa. \tag{8}$$

The first derivative of Eq. (8) with respect to time, using Eq. (3b) gives Eq. (7).
    Note also that Eq. (7) can be written as $\dot{R}/c = (R/ct)(\rho/\rho_2)$ and, since as $R \to \infty$, $\rho/\rho_2 \to 1$, $\dot{R}/c \to R/ct$ which is true in the limit of $\dot{R} = c$. The value of $\kappa$ was suggested by GR-theory, where for radiation domination $G\rho_r t^2 = 3/32\pi$ [8, p 733]. Adopting $\kappa = 3/32\pi$ gives Eq. (1),
    Equation (8) was first derived (1995) from an intuitive expression for cosmic time in terms of the partial times $\Gamma_i^2 = (\kappa/G/\rho_i)$ for each of the three contents of our universe, that add by their inverse squares to give the inverse square of cosmic time $t^{-2} = \sum_i \Gamma_i^{-2}$. This addition immediately produced the beginnings of the SC-theory in terms of the then unknown constant $\kappa$ [3].
    The equivalence of the modified Einstein-Friedmann Eq. (5) and the SC Eq. (1) can be established directly.
    Re-write Eq. (5) in terms of $\rho'$ as,

$$H^2 = (G/\kappa)(2\rho/R\rho')^2 \rho. \tag{9}$$

Take the negative root to get,

$$H = -2(G/\kappa)^{1/2} \left(\rho^{3/2}/R\rho'\right) \rho = (dR/dt)/R. \tag{10a}$$

Using the time derivative of the density in terms of $\rho'$, one sees that Eq. (10a) implies,

$$d\rho/dt = -2(G/\kappa)^{1/2} \rho^{3/2}. \tag{10b}$$

The solution of Eq. (10b) is,

$$t = (\kappa/G\rho)^{1/2}, \tag{11}$$

the square of which is Eq. (8).
    At this point a phenomenological expression for the expansion of 3-D space has been given as Eq. (1) and has been shown to be the solution of a modified Friedmann Eq.

(5). Equation (1) provides an explicit expression for the cosmic time in terms of the 4-D radius R of our 3-sphere universe. So it also determines R implicitly in terms of t.

**3. About the New SC-Theory**

In this paper with the SC-relation Eq. (1) between R and t, cosmic time is $t = (\kappa/G\rho)^{1/2}$. With the scale factor now fixed from the beginning of expansion to far into the future, the description of the *expansion of space* can be considered a succession of discrete states.

In the first part of this paper with no differential equations to solve, calculations were very simple to generate SC-curves by considering all of the APPENDIX equations as one equation. The computer program, with input independent variable of either the ratio of scale factors $R/R_0$ or $Z = (R_0/R) - 1$, sequentially calculated values for all of the equations including time, and for added relations, for every value of input, and only the needed dependent variables were selected for output. A Bisect-Root numerical program [9] was added for input of time as the independent variable for implicit solution of $R$:

$$\rho_0/(t/t_0)^2 = \rho_{r0}(R_0/R)^4 + \rho_{m0}(R_0/R)^3 + \rho_{x0}(R_0/R)^2. \tag{12}$$

**4.0 Predictions of the SC Theory**

Before attacking the SNIa problem some evidence for the good predictions of this theory for expansion parameters is reviewed in this section.

Note in particular that given $\rho_{x0}$ and $\rho_{m0}$, then Eq. (A6) shows that the theory predicts the present density of dark mass ρ$_{x0}$. The predicted current cosmological constants agree with measurements of the Wilkinson Microwave Anisotropy Probe (WMAP) [10]. Thus this SC-expansion phenomenology has no undetermined parameters.. Cosmic time $t(R)$ and the Hubble parameter $H(R)$ in terms of the scale factor R are now known within the range of uncertainty of current measurements of $\rho_{r0}, \rho_{m0}$ and $t_0$ as measured by WMAP.

In a section 5.0 it will be shown that integration over time is required for the transport of radiation inside our 3-D universe [N-D refers to spatial dimensions].

**4.1 Cosmological Parameters**

Values of the present cosmological constants were determined from Eq. (1) according to equations of the APPENDIX. With age set to $t_0 = 13.5$ Gy, the SC-model predicted the following present values for the cosmological constants: $R_0 = 1.354 x 10^{28}$ cm, $H_0 = 68.6$ km s$^{-1}$ Mpc$^{-1}$, $\Omega_B = 0.031$, $\Omega_{DM} = 0.248$, $\Omega_{DM}/\Omega_B = 8.0$, $(\dot{R}/c) = 1.005$ and $q_0 = 0.0084$ (i.e., approaching constant expansion rate $\dot{R} = c$, all within the range of uncertainty of our astronomer's measurements.

Given that the new cosmic time of the expansion is a correct description now and into the future, one might still wonder if it meets the challenge of the very early universe at t of about 10$^{-10}$ s, the sensitive era for ρ(t) of nucleosynthesis. Recently Fig. 6 of Carroll and Kaplinghat [11] shows acceptable model curves of log(H) versus log(a/a$_0$) during nucleosynthesis. The corresponding curve from the SC-theory in terms of $H(R)$ is parallel to the standard model curve (slightly higher) and meets their minimum range of H at $a/a_0 = 8.5 \times 10^{-10}$ with $T_e = 0.277$ Mev.

The log of cosmic time from Eq. (1) is shown in Fig. 1 versus the log of the ratio of scale factors $R/R_0$. The straight-line portion of the curve bends smoothly through matter domination to dark mass domination and to the final straight line into the future. The rate of increase of the 4-D radius of our universe, $\dot{R} = dR/dt$, is always greater than the speed of light c. As $\dot{R}$ continues to **decelerate**, it approaches $\dot{R} = c$ asymptotically, even now and into the future. For $\dot{R}/c$, see Section 2.0.

**4.2 SC vs GR Difference in Predictions**

The new scaling of dark mass $R^{-2}$ implies in large time the limit is $\dot{R} = c$. Even though the SC-theory predicts essentially the same **present** total mass as GR-theory, the change of scaling in theory of SC-dark mass is very pronounced, for the past and future, as shown in Fig. 2 for the difference in total mass content. The predictions are carried into the far future where expansion-$Z \to -1$, i.e., $R/R_0 = 1/(1+Z) \to \infty$. For mass: $M_i(Z) = V_U \rho_i(R) = 2\pi^2 \left( R_0/(1+Z) \right)^3 \rho_{i0} (1+Z)^{Ni}$.

**4.2.1 Expansion Redshift Z** Convention is followed in this paper to indicate the spectral redshift of change in wavelength by (lower-case) $z = (\lambda_{obs} - \lambda_{rest})/\lambda_{rest}$. On the other hand for an emitter at rest in the Hubble flow (peculiar velocity $v_p = 0$ or at large z), the fractional change in wavelength for the cosmological expansion redshift is the same as the fractional change in size of the universe $R$ since the time when the radiation was emitted, $Z = z = (R_{obs} - R_{emitted})/R_{emitted}$ [12]. Astronomers correct their measured redshift z for any motions of source and sink and report the number labeled Z here.

Thus in terms of the present size of the universe $R_0$, when all observations are made, redshift Z is also a measure of the past size $R$ of the universe,

$$Z = (R_0/R) - 1, \qquad (13)$$

using uppercase Z to designate this use. Although z is limited to the past from where we receive radiation, the entire future of our universe occurs over the range $0 \geq Z \geq -1$ and our cosmological predictions must be correct in the future as well as in the past. Use of capital Z allows calculations of many predicted cosmological variables, e.g. $H(Z)$, to extend from the past through Z = 0 toward Z = -1 into the future. Many present concepts could be rejected immediately as unphysical by such simple extension of the calculations.

Of all the literature on the GR-model, the author has not found anywhere that calculations were continued into the future (Z < 0). Of course one cannot measure radiation from the future (spectral z is undefined), but that is independent of how the cosmological parameters change in the future. Perhaps, the next two figures explain the absence of those predictions.

The present and planned great future effort for astronomers to measure the radiation from many past supernova explosions is to establish the very important Hubble parameter H(Z). Equation (1) gives the dimensionless Eq. (7), $tH = \rho/\rho_2$ or the SC-theoretical expression H(Z) simply as Eq. (A16) $H = (\rho/\rho_2)/t$ and $\rho_2 \to \rho$ as $\rho_x$ becomes dominant, or in terms of Z, the SC-$H(Z)$ is,

$$H(Z) = \left(\frac{1}{t_0 \rho_0^{1/2}}\right) \left(\frac{\left[\rho_{r0}(1+Z)^4 + \rho_{m0}(1+Z)^3 + \rho_{x0}(1+Z)^2\right]^{3/2}}{\left[2\rho_{r0}(1+Z)^4 + 3/2\rho_{m0}(1+Z)^3 + \rho_{x0}(1+Z)^2\right]}\right). \quad (14)$$

Clearly the SC-H(Z) is driven to zero in the far future as $Z \to -1$ in Eq. (14) and Fig. 3.

Figure 3 shows the great difference in predictions of H(Z) between the SC-theory and the GR-theory for the GR-case of a flat universe where $\Omega = 1$ at all Z and by forcing $\Omega_\Lambda$ to make up the difference using the equations of Davis and Lineweaver [13] where z is to be replaced by Z,

$$H(z) = H_0(1+z)\left[1 + \Omega_m z + \Omega_r\left((1+z)^2 - 1\right) + \Omega_\Lambda\left((1+z)^{-2} - 1\right)\right]^{1/2}. \quad (15)$$

At Z=0, $H = H_0$ and at $Z = 1$, decreasing Z, decreases $H$, but for $Z \to -1$, $H \to \infty$. The lambda energy drives $\dot{R}$ into acceleration and GR-H toward infinity in the future. To cut off such curves at Z = 0 hides the unphysical prediction of GR.

The GR-dark matter dominated Einstein-de Sitter universe had $tH = 2/3$ whereas SC-$tH \approx 1$. Therefore, for equal $H_0$ in Fig.3, the GR $t_0 \approx (2/3 \text{ SC } t_0) = (2/3 \text{ } 13.5 \text{ Gy}) = 9$ Gy as shown in Fig 4 where the predicted corresponding SC and GR curves for cosmic time are shown.

From the same Eq. (A16) for SC-theory, $t = (1/H)(\rho/\rho_2)$, SC-time in Fig. 4 evolves inversely as H(Z) of Fig.3 or in terms of Z, $t = t_0(\rho_0/\rho(Z))^{1/2}$ or,

$$t(Z) = t_0 \rho_0^{1/2} \left[\rho_0(1+Z)^4 + \rho_m(1+Z)^3 + \rho_x(1+Z)^2\right]^{-1/2}, \quad (16)$$

and clearly $t \to \infty$ as $Z \to -1$. Similarly for GR-theory, Peacock's Eq. (3.44) for k = 0 [14] also predicts $tH \approx 1$ so GR $t \to 1/H$ or $t \to 0$, the opposite as for SC-theory, as $Z \to -1$. The difference between these curves is shown in Fig.4. The GR-prediction for cosmic time could not be considered physical – unless it is cut off at Z = 0.

**4.3 Dark Mass and Rotation Curves**

The main goal is to show that the supernova Ia data support the SC theoretical H(Z) but this new H(Z) depends upon replacement of GR-dark matter with a new dark mass that scales with the expansion as $R^{-2}$. That new scaling must eventually be justified and evidence of its different behavior is thought to be responsible for the flat rotation curves of spiral galaxies and the success in predicting the shape of those rotational curves by the phenomenology of "MOdified Newtonian Dynamics (MOND)"[15]. This author does not accept the basic premise that Newtonian gravity must be modified. Instead the nature of the new dark mass should explain the MOND success. A modified interpretation of MOND has shown how to calculate the equivalent mass of an unknown substance to give the same predictions as MOND [16].

It is easy to show that SC dark mass predicts a flat rotational curve past the optical radius of the spiral galaxy but there is as yet no theory of how the density of dark mass behaves within the optical radius. But it is precisely within the optical radius that MOND does an excellent job of predicting the measured rotational curves. Thus the following is dark mass at the optical radius set equal to the equivalent predicted mass of the modified MOND and then scale as SC-theory predicts at greater radius. The geometry and forces that explain this strange behavior must await a future paper.

Two galaxies were selected to compare the SC-theory predictions with measured velocities. The curves for NGC2903, mass/light (M/L = 1.8), are shown in Fig. 5 and those for NGC1560 (M/L = 1.0) in Fig. 6. The input data $V_T, V_s, V_g$ are shown as symbols where the measured velocities of the orbiting stars $V_T$ are labeled "data" and the estimated contribution from the mass of stars are labeled $V_s$ and that of the gas $V_g$.

Following reference [17], the observed baryonic mass follows Newtonian gravity as $g_N = V_b^2/r$ and all mass follows $a = V^2/r$ by MOND through $\mu(x)a = g_N$ where $\mu(x) = x/\sqrt{1+x^2}$, $x = a/a_0$ and $a_0 = 1.2 \times 10^{-8}$ cm s$^{-2}$. From the mass to light ratio (m/L), with $V_s$ representing the measured velocity of stars and $V_g$ the measured velocity of gas and $V_T$ the total velocity, then let $V_x$ be the unknown velocity of dark mass to be added to the account. Set $V_T^2(r) = V_s^2(r) + V_g^2(r) + V_x^2(r)$. The unknown mass equivalent is $M_x = V_x^2 r/G$. McGaugh's modification (his Eq. (20) to obtain $V_x$ by difference is,

$$V_x^2(r) = \left[\mu^{-1}(x) - 1\right]\left[V_s^2(r) + V_g^2(r)\right]. \tag{17}$$

Thus the distribution of baryons completely sets the distribution of the unknown mass $M_x$ out to the optical radius. In the SC-theory, with no "dark matter", the new "dark mass" scales with the expansion as $R^{-2}$ and, except for gravity, does not interact with ordinary matter. In the 3-D universe, on average, dark mass scales with the expansion as $\rho_x(R) = \rho_{x0} R_0^2/R^2$. This density decreases with the expansion but the total average mass increases as,

$$M_x(R) = \int_0^R 4\pi R^2 \rho_x(R) dR = M_{x0} R/R_0. \tag{18a}$$

Thus small clumps of dark mass should also increase in the same proportion,

$$M_x(r) = M_{x0}\, r/r_0 , \qquad (18b)$$

then,

$$M_x(r) = \dot{M}_x = M_{x0}\, \dot{r}/r_0 = M_{x0}\left(\dot{r}/r\right) r/r_0 = M_x(r) H . \qquad (19)$$

The Hubble parameter H is very large in the past, so even the very small early clumps of dark mass increase very rapidly and could be expected to generate early black holes and seeds for condensation of baryonic matter.

From Eq. (17) at the optical radius, then $M_x = V_x^2 r_{opt}/G$ and increases thereafter as $M_{x0}\, r/r_{opt}$ to produce a constant $V_x$ for $r > r_{opt}$ and an asymptotic flat rotation curve for $r \gg$ optical radius.

To show that the SC-theory with new dark mass does indeed, successfully predict rotation curves, the data for the rotation curves of two different galaxies (NGC2903, NGC1560) were obtained from S. McGaugh [17]. A computer program was written following the procedure to extract the equivalent SC-dark mass (x = DM):

$$V_T^2/r = \left(V_b^2/r\right)\left(\left(1+(a/a_0)^2\right)^{1/2}\big/ a/a_0\right) \text{ where } V_b = \left(V_s^2 + V_g^2\right)^{1/2} \qquad (20)$$

The contributions to the total velocity $V_T$ are $V_s$ due to stars, $V_g$ due to gas and $V_x = \sqrt{V_T^2 - V_b^2}$ due to dark mass.

After the optical radius, the total mass of stars and gas for $r < r_{opt}$ were combined $M_b$ (b = baryons) and the equivalent velocity $V_b = \sqrt{GM_b/r}$ gives their Kepler-type decline with increasing r, which makes $V_T$ decline and approach $V_x$.

NGC2903 of Fig. 5 is an example of an early-type disk galaxy where the rotation curves rise rapidly to high rotation velocities of order 200 km s$^{-1}$ with the central region dominated by stellar mass [18]. NGC1560 of Fig. 6 represents later-type galaxies where the reproducing dark mass has become the dominant mass in the central region.

**5.0 SC-Transport of Radiation in An Expanding Universe**

To really test whether a theory has the correct Hubble parameter over time H(t), measurements are needed of radiation that has traveled from enormous distances in the past. Just such measurements have been made of radiation from a special class of exploding stars called supernova Ia. Many of the SNIa exploding stars are found nearby but few are found out past redshift of expansion of Z = 1.7 when our 3-D universe was only a fraction 1/2.7 of its present size, $R_0 = 4388.$ Mpc.

The SC-theory as embodied in Eq. (1) has no need for dark energy and does not predict accelerated expansion, as current theory does to predict the supernova Ia measurements. So far SC-theory says nothing about the transport of radiation. The following discussion shows that the SC-theory can successfully account for these supernova Ia measurements without the added burden of dark energy. It does this precisely because it has the better evolution of the scale factor $R(t)$ contained in its Hubble parameter $H(Z)$. GR-transport of radiation will be summarized and compared at the end of this Section 5.

Astronomers speak of exploding SNIa stars as "standard candles" in that they seem to have nearly the same maximum luminosity $L$. From the well-known formula for the measured flux $F$ of radiation from local sources, $F = L/4\pi r^2$, a distance $r$ can be inferred from the measured $F$ and luminosity $L$. But in cosmology the corrections for the expansion and relativistic effects for the luminosity distance $r \to d_L$ are so enormous that *distance* looses its usual meaning as a time-of-flight measurement between the end points.

With our universe as an expanding 3-sphere, $R$ is its radius in units of Mpc (3.0856x10$^{24}$ cm) and our new cosmic time of Eq, (1) in units of Gy (billion years) has simultaneous time $t$ everywhere at size $Z$ and the local speed of light is c = 306.6 Mpc/Gy. Consider two objects a and b at rest in the expanding universe, peculiar velocities $v_p$ = 0, at a distance $r$ apart at time $t$. The Hubble law says each measures the other expanding away at radial Hubble velocity $v_H = Hr$.

**5.1 Graphic Illustration of an Expanding 3-Sphere**

A second assumption in the Introduction to this paper was that our universe is a closed 3-sphere. The graphic illustration Fig. 7 provides the path of radiation in our 3-D space as that 3-D space expands radially in the fourth spatial dimension.

Three concentric circles of increasing radius $R$ in Fig. 7 represent one dimension of the expansion of our 3-sphere universe of three spatial dimensions. There is no fourth spatial dimension in GR-theory.

At any one time, only one circle exists with points on it where the packet of photons being followed exist and the points where the radial world lines of stationary sources cross that circle. The largest circle, Z = 0, represents our present universe of radius 4388 Mpc and age 13.5 Gy. The spirals represent the path of radiation trajectories and for now we are concerned only with the smaller trajectory 1 that will end at point "a" inside our astronomer's telescope. The telescope is pointing clockwise at point "d", the region of space containing the remains from a SNIa that exploded some 8.9 Gy ago, at point "e" and whose radiation followed the spiral trajectory to reach the telescope at the present.

Light in the vacuum (radiation) must always travel at velocity c in our 3-D universe represented by the circle even as that circle expands $(\dot{R} \geq c)$ always normal to the direction of the light. From Eq. (27) to be derived next, trajectory 1 for the cosmic microwave background (CMB) radiation was started counter-clockwise shortly after the decoupling of radiation and matter at Z = 999, $t = 6.08 x 10^{-4}$ Gy at a clockwise fraction f = 0.65517 of the circumference then, or at an emission distance $r$ = ED = 18.06 Mpc from the astronomer's world line. Those CMB photons were only 58.9 million light years (Ly) away as they headed directly toward the position in space where they would end up, but it took them (13.5 – 0.00061) billion years to get there. All points on trajectory 1 are potential emission distances for the astronomer.

The emission distance is an interesting concept in that those CMB photons were emitted even before our solar system was formed. So from what marker is this emission distance really considered? It is assumed that some of the CMB hydrogen atoms from back then, also made it all of the way to the present and into the frame of the telescope (or astronomer) and their radial world line is the vertical line from the origin o to point a. The 3-D emission distance r = ED is always measured on the arc of the circle from the current light trajectory to the world line o – a.

Thus there was a fourth circle around the origin (not visible) of radius $R_{em} = 4.38$ Mpc at $R_0/R = (1+Z) = 1000$ and the photon source, at a point on the circumference (from a, angle $\theta = 2\pi f$), that produced the CMB source whose radial world line expands and ends at point c at the present.

The circle labeled Z = 1.7 is a fascinating feature of this 3-sphere plot of the SC-theory. It turns out in the SC-theory that the arc distance ED at Z = 1.7 is always maximum, and this is so in the near past and into the future. Those that studied the Einstein-de Sitter universe may remember that its maximum was at Z = 1.25 [19].

This SC Z = 1.7 feature is now combined in SC-theory with another feature that the (local) compounded (approaching) velocity of light (vacuum) **in an expanding universe** is $v_c$. Consider again the two objects, a and b, at rest in our expanding universe and place the origin of the reference frame at a. Radiation is approaching in the space at b at velocity –c but the Hubble law says 3-D space itself at b is receding at the Hubble velocity $v_H = Hr$. Therefore the net velocity of the radiation at b toward a, for a is,

$$v_c = dr/dt = Hr - c. \tag{21}$$

At any one cosmic time, this simple compounded velocity of radiation combines the two basic perpendicular velocities: the 3-D speed of light c and the 4-D radial expansion rate $\dot{R}$ of $H = \dot{R}/R$ where $\dot{R} \geq c$. Although Eq. (21) may not be consistent with the special theory of relativity [21], neither does the special theory take any account of the expansion of our universe.

Equation (21) is the net velocity of radiation at distance r as it approaches the past world line of the atoms of the astronomer's telescope and we need its solution for the emission distance ED(Z) at redshift Z = z for later use in the astronomers luminosity distance $d_L$ expression.

An approximate analytic solution of Eq. (21) was obtained with the integrating factor

$$\alpha(t) = \exp\left[\int_{t_e}^{t} H(t')dt'\right], \tag{22}$$

as now outlined. A common approximation in cosmology at low redshift is to neglect the very low energy density of the CMB microwave radiation. Here the interest is in the transport of much higher energetic visible and infrared photons with negligible interaction with the microwave photons. Thus the major density terms of matter and dark mass can be expressed in the dimensionless ratio $\varsigma = \rho_{x0}/\rho_{m0}$ and Z. From Eq. (1)

$t^2/t_0^2 = \rho_0/\rho$ and its derivative $dt$ and $tH = \rho/\rho_2$ gives $Hdt = -1/2(d\rho/\rho_2)$. From Eqs. (A10 to A12), $d\rho/\rho_2 = 2\,dZ/(1+Z)$, so $Hdt = -dZ/(1+Z)$ and,

$$\alpha = \exp\left[\ln\left((1+Z_e)/(1+Z)\right)\right] = (1+Z_e)/(1+Z). \tag{23}$$

Let,

$$X(Z) = \alpha\, X(Z_e) - \alpha c t_0 \int_{t_e}^{t_0} \alpha^{-1}\, dt'/t_0. \tag{24}$$

From $t/t_0 = (\rho_0/\rho)^{1/2}$ after differentiation,

$$\frac{dt}{t_0} = -\frac{3}{2}\frac{(1+\varsigma)^{1/2}}{(1+Z)^2}\frac{\left[(1+Z)+2\varsigma/3\right]}{\left[(1+Z)+\varsigma\right]^{3/2}} dZ. \tag{25}$$

Substituting $dt/t_0$ from Eq. (25) into the last integral of Eq. (24) leads to a number of standard integrals involving logarithms. Now define a function,

$$U(\varsigma, Z) = \left(\frac{1}{\sqrt{\varsigma}} \ln\left[\frac{\sqrt{\varsigma+(1+Z)} - \sqrt{\varsigma}}{\sqrt{\varsigma+(1+Z)} + \sqrt{\varsigma}}\right] - \frac{1}{\sqrt{\varsigma+(1+Z)}}\right). \tag{26}$$

After successive integrations solve for $ED = X(Z_e)$ with $ct_0 \approx c/H_0$ and generalizing $Z_e$ for any $Z$,

$$ED(Z) = \left(\frac{c}{H_0}\right)\frac{(1+\varsigma)^{1/2}}{(1+Z)}\left[U(\varsigma, Z) - U(\varsigma, Z=0)\right]. \tag{27}$$

This is the (approximate) equation for the spiral trajectory with $X = R\sin(\theta)$, $Y = R\cos(\theta)$ and $\theta = r/R$. With the telescope pointed in a fixed orientation in space, all photons that enter have traveled on the same spiral trajectory that is also the $ED(Z)$ emission curve of Eq. (27). For large Z, a complete solution of Eq. (21) is needed.

Thus at $Z = 1.7$, $v_c = 0$ and the net velocity of radiation toward the astronomer is momentarily zero. Inspection of the CMB trajectory of Fig. 7, shows that for the 4.58 Gy before reaching $Z = 1.7$, the CMB photons were actually moving further away from the astronomer's world line. Such behavior is clearly shown in Fig. 7 using Eq, (27) with the $\rho - t$ expression of Eq. (1).

The implication of this feature for the astronomer is that during that long period of negative approach, the expansion was rapidly diluting the energy density of the radiation from both classical and relativistic effects. It can be expected that measurements of radiation from sources that approach and exceed Z = 1.7 will be very sensitive to errors in measurement of their spectral redshift z.

Another interesting feature of $v_c = 0$ at Z = 1.7 comes form the SC-prediction that now and into the future, $\dot{R} \approx c$. Substituting $\dot{R} \approx c$ into Eq. (21) for $\dot{R}$ of $H = \dot{R}/R$, gives $r \approx R$, or in other words at Z = 1.7, the emission distance approximates the radius of the universe

At present, the maximum Z of observed supernova Ia is also about Z = 1.7 and that one SN1997ff even had a lensing assist. Suppose there was a supernova Ia explosion at Z = 1.7 at point e. The debris of the source would have a radial world line from e to point d. Its ED = 1595 Mpc is only 5.20 G ly (billion light years) from the astronomers world line. Nevertheless, because those photons start toward the telescope with zero net velocity, it takes them 13.5-4.58 = 8.92 Gy to reach the telescope. Therefore the final decreased flux of radiation measured by the astronomer converts to a far greater "distance" than the ED = 1595 Mpc 3-D distance when the photons began their travel.

If the remains of the star do not acquire a peculiar velocity, $v_p = 0$, then the expansion carries the remains from point e to point d at the present, which is reception distance RD of the source and from Eq. (27) for $ED(Z)$,

$$RD = (R_0/R)ED = (1+Z)ED, \qquad (28)$$

where $ED(Z)$ is given by Eq. (27).

## 5.2 Theoretical Check on SC- $H(Z)$ Space-Time Predictions

To verify the global predictions of the new SC-Hubble parameter $H(Z)$, the goal is to compare its predictions to the measurements of supernova Ia radiation from the past. But is there an existing theoretical prediction that can be used to test SC- $H(Z)$ over an extended period of time. Consider trajectory 2 in Fig. 7. A decade or so ago, astronomers hoped to determine the cosmological constants with measurements of the deceleration q of the universe. The favorite Einstein-de Sitter universe of the Friedmann equations had a q $\approx 1/2$ (unlike the SC-q $\approx 0$). A. Loeb in such a study of the cosmic deceleration [22], mentions Weinberg's equation [23],

$$\int_{t_s}^{t_0} dt/a(t) = \int_{t_s+\Delta t_s}^{t_0+\Delta t_0} dt/a(t), \qquad (29)$$

where $a(t) = (1+Z)^{-1}$ and yields $\Delta t_s = \Delta t_0/(1+Z_s)$ for the relation between time intervals of emission $\Delta t_s$ at $Z_s$ and observation $\Delta t_0$ at Z = 0.

This relation may be adequate for time intervals of order centuries but how does it fare for long periods of order Gy where time varies as in Eq. (1)? The second trajectory 2 of Fig. 7 was calculated to provide an answer. The emission time for trajectory 1 was $t_s = 0.0006$ Gy at Z = 999, and ED = 18.0693 Mpc at f = 0.655171 in Section 5.1.

An observation interval of $\Delta t_0 = 1.5$ Gy was selected for a future universe of age $t'_0 = 13.5 + 1.5 = 15.0$ Gy and radius $R'_0 = 4849$ Mpc. An input of $t'_0 = 15.0$ Gy to Eq. (1) promptly returns the implicit output of $R'_0/R_0 = 1.105$. In terms of the present universe, the future universe is at $R'_0/R_0 = 1/(1+Z) = 4849/4388$ or Z = -0.0950 and densities and other parameters are readily calculated.

The general formula for translation of present values of Z to new values Z' from the new basis $R'_0$ is

$$Z' = (R'_0/R_0)(1+Z) - 1. \qquad (30)$$

For our present Z = 0 universe, $Z' = 0.105$ and our special $v_c = 0$ at Z = 1.7, gives $Z' = 1.98$. The formula also applies for translation to an earlier universe $R'_0/R_0 < 1$.

From the basis of this future universe, also with the CMB photons starting from the same fraction of the circumference at $f' = 0.65517$ as in Section 5.1, it was found from Eq. (27) that emission at $Z'_s = 483.0$ at $t'_s = 0.0028$ Gy and ED' = 41.251 Mpc resulted in the new photons arriving on time, $t'_0 = 15.0$ Gy at Z' = 0 and $f' = 0$, i.e., trajectory 2.

The difference in emission times of trajectories 1 and 2 is $\Delta t_s = 2.8 - 0.6 = 2.2$ million years (My) but we have two different expansion redshift $Z_{em}$. Trajectory 1 where $(1 + Z_{em}) = 1000$, predicts $\Delta t_s = 1.5$ Gy/1000. = 1.5 My, whereas trajectory 2, with $(1+Z'_{em}) = 484.$ predicts $\Delta t'_s = 1.5$ Gy/484. = 3.1 My. The average, $\langle \Delta t_s \rangle = 2.3$ My is in fair agreement with the predicted $\Delta t_0 = 2.2$ My of Eq. (29).

**5.3 Additional Features Predicted by a 3-Sphere Geometry**

Other features of radiation trajectories are shown in Fig. 8. If the telescope was pointed in exactly the opposite direction, the left trajectory 2 represents all of the sources of radiation that could enter the telescope at present. Symmetry predicts its CMB source would follow a different radial world line o – g in a different part of the universe.

But what of the original source of trajectory 1 photons that were emitted in the opposite direction? Again symmetry predicts a new very different trajectory (not shown but call it 3 of the same shape but rotated by angle $\pi$ about straight-line c - o – d, which is passing through point e and ends at point h.

Consider again point e. If the photons of both trajectories 1 and 3 from the same patch of CMB space pass through point e, then every photon emitted radially from that patch has the potential to pass through point e. Our 3-sphere universe re-focuses radiation (expanded virtual source) at the antipode of the source. An expansion redshift of $Z_{em} = 999$ is before the evolution of the large sources of radiation and it re-focused at a fairly large $Z$ of 1.7, but that means that the CMB radiation had already refocused at t = 4.58 Gy, some 8.9 Gy ago. Thus more evolved patches of our past universe are refocusing now on our present universe. Note X-ray sources will produce infrared antipodes and if past high energy sources just happen to pass through our antipode the increased infrared radiation on our planet might be severe for a few hundred years.

Possible sources of radiation that are focused on our present universe at point a are clearly demonstrated in Fig. 8 at its antipode, just below the origin where trajectories 1 and 2 first crossed. That source is at f = 0.5, Z =35.3, t = 160 My and ED = 380 Mpc when $R$ = 121 Mpc. This re-focusing of radiant energy results from the major difference between the GR-flat and SC-closed geometries. The non-focused early X-rays may account for a fraction of the present excess background infrared radiation [24] in contrast to the current big-bang infinite universe where re-focusing is not possible.

Figure 8 predicts another interesting phenomenon, which has the potential to be measured by a future instrument, called an "Anoka scope". Consider the point marked f where trajectory 2 crosses circle Z = 9. If an instrument in orbit has been built to record the pattern of the CMB at one end and the pattern of Z = 9 sources at the other end, then it should be measuring the very same patch when it emitted the CMB and the mirror image of the Z = 9 evolved patch – independent of its orientation in 3-D space..

**5.4 Effects of Accelerated Expansion**

To check the effects of an added acceleration to our expanding but decelerating universe a special calculation was made. The same calculation using Eq. (22) for trajectory 1 in Fig. 7 was repeated except that the Hubble parameter was held constant, after the photon packet reached $Z = Z_K$, to $H(Z_K) \to \dot{R}/R = H_K$. Radius $R$ continues to increase, so $\dot{R}$ must increase and thus $R$ must accelerate. At $Z_K = 1.8$, $t_K = 4.4$ Gy, $H(Z)$ reached the constant $H_K = 200.$ km s$^{-1}$ Mpc$^{-1}$ was then held constant and the trajectory in Fig. 8 curved out radially past point "i", None of these photons could ever reach world line a. This quick calculation was not adjusted to account for a larger $R_0$ with the acceleration.

To predict what the astronomer would observe because of the hypothetical acceleration, keep the telescope at the top and rotate new trajectory 3 and its source world line about 30 degrees counter-clockwise to end the new trajectory at point a.. The astronomer still views the CMB of the same age, but from a less distant patch. The change in trajectory for acceleration of constant rH = c at Z = 1.7 could also be calculated.

**5.5 GR-Transport of Radiation**

Hubble's measurements of galactic distances quickly convinced Einstein that our universe was expanding and Einstein quickly removed the lambda constant $\Lambda$ he had added to his GR-theory to force a static universe. Now GR-theoreticians have added it back with opposite sign to force a fit to SNIa data, which also forced an acceleration of the expansion rate $H$ as shown in Fig. 3.

The amount of lambda that is added is most often just that to force the total density parameter $\Omega = 1$, to justify the period of inflation added to solve their "beginning" GR uniformity problems. However, in turn, $\Omega = 1$ means in GR-theory that our expanding 3-D universe is infinite. But unlike SC-theory, GR-theory has no outside higher-dimensional space for even a closed universe to expand into.

Another notion in Einstein's GR-theory is that any one coordinate reference frame is as good as another as long as it is compatible with his field equations. The Friedmann equations provide such a coordinate system but they present an expanding distance as the product of the distance now from the Earth $\varpi$ (comoving distance) [12] times a changing ratio of dimensionless scale factors $R(t)/R_0 = 1/(1+z)$. Of course, $R_0$ is undefined in an infinite universe, so $R_0$ is set to unity and $a = R/R_0 = R$.

Following Reference [12], the proper distance from the Earth toward an object in terms of the Robertson-Walker metric with ds = 0 and $(dt = d\theta = d\phi = 0)$ is,

$$d(t) = R(t)\int_0^\varpi \left(1 - k\varpi'^2\right)^{-2} d\varpi' = R(t)\varpi, \qquad (31a)$$

for an infinite universe $k = 0$.

For a traveling photon from emission toward Earth for a flat universe $(k = 0, \Omega = 1)$, the distance to that photon is,

$$d(t) = R(t_0')\varpi = (2c/H_0)\left(1 - (1+z)^{-1/2}\right). \qquad (31b)$$

The radiation from a nearby source of luminosity L arrives at radius r with flux $F = L/4\pi r^2$. Thus if the corrections for the expansion are put on a correction of $r \to dL$, including the two (1+z) corrections for the relativistic diminution of $L$, one obtains for the luminosity distance $d_L^2 = L/4\pi F$ or

$$F = L\big/\left[4\pi\varpi^2(1+z)^2\right] \text{ or } d_L = \varpi(1+z). \qquad (32)$$

In terms of the expansion constants for $\varpi$, Eq. (32) leads to the $d_L$ equations [12]:

$$d_L = \frac{2c}{H_0}\frac{1}{\Omega_0^2}\left[\Omega_0 z - (2 - \Omega_0)\left(\sqrt{1 + \Omega_0 z} - 1\right)\right], \qquad (33a)$$

$$d_L = \frac{c}{H_0}\frac{1}{q_0^2}\left[q_0 z - (1 - q_0)\left(\sqrt{1 + 2q_0 z} - 1\right)\right] \qquad (33b)$$

For the last Eq. (33b), SC-theory predicts $q_0 \approx 0$ now and into the future. However, Eq. (33a) offers an opportunity later to check the SC-equation for the luminosity distance $d_L$ to be derived next from Eqs. (1) and (27).

**6.0 Supernova Ia**

Measurement of the radiation from the class of exploding stars called: Supernova Ia is important because they are considered to explode with about the same maximum luminosity L. Astrophysicists explain the exploding star is a member of a binary that has already burned its own hydrogen down to carbon and is in the process of capturing more mass from its expanding companion. When the captured mass reaches a critical value, the carbon burning is triggered and the total star explodes. The explosion is so bright that astronomers can capture their light curve back to about z = 1.7.

**6.1 SC-Prediction of Luminosity Measurements of Supernova Ia**

From the SC-point of view, radiation is confined to move in 3-D space. But to repeat, it also moves in the radial direction, not because the vector $\vec{c}$ has any component in the radial $R$ direction, but because our 3-D space itself is expanded in the radial direction $\vec{R}$ as shown in Fig. 7. So radiation moves on great circles of the expanding 3-sphere of circumference $C = 2\pi R$. Given the expansion redshift $Z$ of the supernova Ia, the SC-expansion theory provides the time of emission $t_e$. If there were no expansion, the speed of light c would expand its 2-sphere pulse of SNIa radiation to the present, a radial distance of $r = (t_0 - t_e)c$ and as in Section 5.5, the flux F at $r$ would be $F = L/(4\pi r^2)$ where $L$ is the luminosity of the source.

But during that interval of time, the expansion has also decreased flux by expansion in the radial direction equivalent to the distance it moved the star remains according to Eq. (28), which is (RD – ED) = ((1 + Z)·ED – ED) = Z·ED. The equivalent total is,

$$d_4 = c(t_0 - t_e) + Z \cdot ED, \tag{34}$$

where $d_4$ would replace radius r in the flux equation. However, there are two relativistic effects that each decrease $L$ one factor of $(1+Z)$ due to the decrease in photon energy and another $(1+Z)$ for the dilation of time interval between photons, the same as for the BB-theory, $L' = L/(1+Z)^2$. But $d_4$ is squared; so only one factor of $(1+Z)$ is used for the luminosity distance $d_L$,

$$F = L/4\pi (d_4 \cdot (1+Z))^2 = L/4\pi d_L^2. \tag{35}$$

The SC-luminosity distance is,

$$d_L = d_4(1+Z) = (c(t_0 - t_e) + Z \cdot ED(Z))(1+Z). \tag{36}$$

The emission distance ED(Z) of Eq. (27) is obtained by an analytical solution obtained in good approximation by neglecting radiation and expressing matter and dark mass as the ratio $\varsigma = \rho_x/\rho_m$ [20]. $ED(Z)$ increases with increasing $Z$ until $Z = 1.7$ and then it decreases with increasing $Z$. In units of $c/H_0$, Eq. (36) becomes the dimensionless expression,

$$\frac{d_L}{(c/H_0)} = \left(\tfrac{2}{3}\right)\left(\frac{(1+\varsigma_0)}{(1+2\varsigma_0/3)}\right)\left[\left(\frac{1}{((1+Z)+\varsigma_0)^{1/2}}\right) + \frac{Z \cdot ED}{(c/H_0)}\right](1+Z). \quad (37)$$

Note that combining Eq. (27) for $ED(Z)$ with Eq. (37) gives for a known luminosity $L$, the implicit dependence $Z(F)$ for a measured flux F. That is, with theoretical $H(Z)$ known, the emission redshift $Z_e$ could be easily back calculated with a computer if the effective measured flux $F$ was reported. This powerful claim applies to any "uniform radial" source of radiation; e.g., SNIa or GRB. The comparison of the SC-$d_L$ of Eq. (37) and the GR-$d_L$ of Eq. (31a) is shown in Fig. 9 for the same value of $\Omega = 0.278$. Although the approach to derivation is very different, the major difference is that between the SC-scaling of $R^{-2}$ for dark mass and GR-scaling of $R^{-3}$ for dark matter.

### 6.2 Astronomer's Scale of Brightness

Hipparchus began, and astronomers have continued, their own relative numerical scale of "apparent magnitude (m, mag)" of the "brightness" of stars in the sky following early astronomer's use of a power scale. [12].

Relative to our Sun of measured distance r = 1 AU = 4.848x10$^{-6}$ pc and measured luminosity $L_\odot = 3.826x10^{33}$ ergs s$^{-1}$, one can convert the relations for **apparent magnitude m** to one in terms **luminosity** of a source $L$, ergs s$^{-1}$ and its effective **luminosity distance** $d_L$, pc (1 pc = 3.0856x10$^{18}$ cm), but placed at 10 pc,

$$m = 4.76 - 2.5\log(L/L_\odot) + 5\log(d_L/10), \quad (38)$$

or for the Sun, $m_\odot$ = -26.81 mag. The first two terms on the right give the absolute magnitude M of the source, (The second term is zero for the Sun.). First note that an **increase** in luminosity $L$ of the source **lowers** the apparent magnitude m. Moving the absolute magnitude M to the left of the equal sign in Eq. (38) gives in general,

$$m - M = 5 \log (d_L/10 \text{ pc}), \quad (39)$$

and $m - M$ is called the **distance modulus**. The apparent magnitude $m = (m - M) + M$ and M is the absolute magnitude of the SNIa source.

The astronomer actually measures the focused flux F of incoming radiation, Eq. (35), by photon counters, ergs s$^{-1}$ cm$^{-2}$. He does not report that information directly but instead assumes a cosmological theory and reports either the distance modulus m-M or apparent magnitude m of the source from the known M of (nearby) supernova Ia and the calculated luminosity distance $d_L$, using the measured spectral redshift z of the host galaxy. Thus it should be instructive to compare the SC-$d_L$ predictions of Eqs. (37) and the GR-$d_L$ predictions of Eq. (31a) where in Fig. 9 both have the same $H_0 = 68.6$ km s$^{-1}$ Mpc$^{-1}$ and total mass $\Omega_0 = 0.278$.

Assuming the SC-$d_L$ curve in Fig. 9 is correct, it can already be seen why the GR-theorists were led to add their additional dark energy to their incorrect cosmology. In the critical range $0.5 \leq Z \leq 2.0$ of Fig. 9, assuming the measured flux follows the SC-$d_L$, the low GR-$d_L$ predicts a low distance modulus m-M. A GR-accelerant $\ddot{R}$ was needed, so the accelerant $\Omega_\Lambda$ was added as in Eq. (15), which in this critical range did decrease H and increase $d_L$ and m-M. But as was shown in Figs 3, such an accelerant also sends $H \to \infty$ in the future and so produces non-physical prediction of both $H(Z)$ an $t(Z)$ in the future.

Besides the comparison of luminosity distance $d_L$ for the two theories, the other two curves in Fig. 9 show the magnitude of the two components that sum to SC-$d_L$ with the speed of light dominant below $Z \approx 1.7$ and expansion above $Z \approx 1.7$. On the question of SNIa as standard candles and the luminosity distance $d_L$ as a real distance in our universe, note from Fig. 9 for a source at $Z = 1.7$ that $d_L \sim 15{,}000$ Mpc or a factor 9.4 greater than the real distance ED and a factor 3.5 greater than the real distance RD, So $d_L$ is an adjusted parameter with units of distance to account for the expansion and relativistic effects. However, as derived in SC-theory, $d_L$ would be equivalent to the real distance to the SNIa source in a larger static 3-sphere universe with no relativist effects.

**6.3 Supernova Data**

As the author understands the measurement, the flux of focused photons to the detector at one of a number of wave bands are measured periodically for a number of days to get the maximum magnitude, after which periodic measurements are made over a number of weeks to fix the decay rate. After years of such measurements, the astronomers have learned to make adjustments for variations from the norm.

The other key measurement of the spectral redshift z of the host galaxy is taken after luminosity decay of the exploding star and often with a different telescope.

The flux of photons, or of total energy, is not reported, but a cosmological model is assumed and the collected information converted and reported versus **redshift z**, either as the **distance modulus m-M** or the **apparent magnitude m** where units of **"**mag" are often given. Sometimes other symbols are used for small corrections but a quick view of the data will generally reveal which magnitude it is. The relation is m = (m–M) + M where M, the absolute magnitude. From an average of some nearby supernova Ia data, M was set to -19.34 [32]. A plot of (m-M) vs redshift Z is called a Hubble diagram. A. V. Fillippenko [25] and others have written much on these supernova Ia measurements.

The comparison of SC-theory and measurements are shown in Fig. 10 for a recent collection called the "Gold and Silver Data" by Riess, et al. [26]. The fit is good but the large scatter of the data precludes distinguishing the correct theory.

**6.4 Analysis In the $H/H_0 - R/R_0$ Plane**

Analysis of such data was greatly improved by a procedure recently developed by Padmanabhan and Choudhury [27]. In this procedure one simply uses the test cosmological theory to calculate for each data point value of apparent magnitude m (or distance modulus m–M) at $Z$, the corresponding value of $H(Z,m)$ for that $Z$ and then present the data in a new plane of $H/H_0$ versus $R/R_0 = 1/(1+Z)$ together with the corresponding theoretical curve $H(Z)$.

Their presentation for the big bang theory has been reproduced in Fig. 11 with the kind permission of joint author T. R. Choudhury. The scatter of the data is greatly increased. Note that the trend of the measured data points fit **NONE** of the versions of the big bang theory. That fact indicates the GR-cosmology is incomplete.

Application of the procedure to the Gold and Silver data of Riess, et al. is shown in Fig. 12 and here the trend of the data does indeed follow the SC-theoretical curve. Likewise the scatter of the data is greatly increased because of the greater sensitivity in the past to the fundamental rate of expansion $H(Z)$ instead of the size of the universe Z as was shown in Fig. 3.

Calculations of the theoretical curve were also made with the value of Z offset $\delta Z \pm 0.1$ and these two curves appear to predict the bounds of the increasing scatter of the data with increasing Z. This procedure was applied to other sets of SNIa data with similar results [20]. For further reference call this: "the **HZ-Process**".

If the SC-theory is correct, as indicated in Fig. 12, it suggests that some information might be gained by forcing the data to fit the theoretical curve. There is no guide for how to move the points in the raw data of Fig. 10 to force the data points to be consistent both with $H$ and $Z$ on the m(z) theoretical curve. The $H/H_0$ vs $R/R_0$ plot of Fig 12 does offer such a guide. Forcing a fit in the $H/H_0 - R/R_0$ plane of Fig. 12 would guarantee consistency with SC-$H(Z)$ theory at m (apart from accuracy of m).

Assuming the major error is one of redshift Z in Fig. 12 for the Gold and Silver data, the measured m remains the same, and a fit of $H(Z,m)$ to $H(R(Z))$ of Eq. (A15) of SC-theory was forced in the $H - R$ plane of Fig. 12. Because of the sensitivity of $H(Z,m)$ to m, an iterative calculation was used to find the new value of Z' such that its $H(Z',m)/H_0$ was within a tolerance $\phi$ of the SC-theoretical value (curve of Fig. 12),

$$\Delta Z = 0.005 \cdot Sgn \cdot Z \cdot \left| \left(H(Z,m)/H_0 - \left(H(Z)/H_0\right)\right) \right|, \qquad (40)$$

where Sgn is the sign of the bracketed quantity. After each $\Delta H/H_0 > \phi$ correction, a new $\Delta Z$ was calculated for the same m and the process repeated until the difference was less than $\phi = 0.005$. All of the data points did converge to the theoretical curve within the tolerance of $\phi = 0.005$.

The 185 values of the new final $\Delta Z = Z' - Z$ were saved. The average deviation from the theoretical curve was $<\Delta Z> = -0.0027$. The $\Delta Z$ values were then applied to the original m data of Fig. 10 and all of the points fell on the theoretical curve of m vs Z. The Hubble parameter value $H(Z)$ for each point then agrees with theory that was not the case for Fig. 10.

Such manipulation of the raw data cannot determine the source of errors in the data, but it does suggest the theory is correct. The forced fit of data to theory could hardly be accomplished in the big-bang model with the theoretical uncertainty shown in Fig.11 since none of the seven models shows a fit to the data. The forced fits are not shown here but will be included in a different type summary plot later in the paper. With added forced fit, for further reference, call this "the "HZF Process" The HZ-process has theoretical support, the HZF-process does not.

**6.5 Spectral z versus Expansion Z**

If the astronomer's appropriate flux F could be reported along with their cosmological-fit magnitude m, then one could calculate an expansion Z from the flux to compare to their measured spectral z. Although the expansion Z would be contaminated with the present GR dark energy model, the principle can be demonstrated by plotting the measured spectral z versus the Z determined from the "HZ process" in the $H/H_0 - R/R_0$ plane. The results of the spectral z, versus expansion $Z$ analysis are shown in Fig. 13 for the Gold and silver Riess data and the recent SNIa data of Astier, et al. [28]. The fair agreement of z with $Z$ shows that the GR added dark matter and dark energy managed to produce a fair fit of m(z) to GR-Friedmann theory as in Fig. 10.

**6.6 Revised SNIa**

Revised SNIa values have been published. One such revised set has been analyzed in terms of the SC-theory. Of the above 185 Gold and Silver SNIa of Riess, et al. [26], Riess, et al. [29] selected 23 SNIa to recalculate the value of $\mu_0^a$ (m-M). Of these, 20 were selected of which 19 appeared on the last page of the list of the former 185 SNIa. Essentially only the values of m-M were upgraded. Also Riess, et al [29] added 21 new high-z SNIa, also labeled of "Silver and Gold" quality. Let $\dot{a}_i = (H/H_0)_i$

The results of the above HZ-process on both sets of data are shown in Fig. 14. As with previous sets of SNIa data, the scatter of the data increases rapidly with increasing Z. The average deviation from the SC-theoretical value $\mu$, $\alpha = \sum_i |\dot{a}_i - \mu|/N$ of the N = 19 SNIa from Fig 14 was $\alpha = 0.3771$ while that of the same revised "old" gold and silver SNIa was $\alpha = 0.3478$ –a slight improvement. The new data was slightly higher with $\alpha = 0.4009$. Note that the error in the Hubble parameter $H(Z)$ is of the order 30 – 40 % at high Z for the new data, nevertheless, the average of the mean deviation $\bar{d} = \sum_i (\dot{a}_i - \mu)/N$ was only $\bar{d}$ = -0.00103 indicating the same trend of SC-theory. For three of the high-Z SNIa in Fig. 14, the forced-fit, HZF process was captured by the computer from the symbol shown to the theoretical curve (no final symbol). From above, increasing Z (decrease of a) produces a decrease in $H$ toward SC-$H$, and from below, lowering Z (increase of a) produces an increase in $H$ toward SC-$H$.

These data confirm again the weakness of the m(z) plot like Fig. 10 to judge the quality of the data vs theory. To test a cosmology for its prediction of the scale factor $R(t)$ or Hubble parameter $H(z)$, one must extract $H(Z)$ and compare to the theoretical prediction of $H(Z)$ as in Fig. 12 or 14 and not as in Fig. 10.

**7.0 Gamma-Ray Bursts**

Recently, B. F. Schafer [30] extended the Hubble diagram $H(z)$ to redshift Z = 6.6 with 69 gamma-ray bursts (GRBs). His analysis includes for dark energy, both the present BB-concordance model with constant cosmological constant and the Riess model (above "Gold data") with varying dark energy. Here the interest is in the raw data and whether it can be fit by the SC-theory that has no dark energy.

There are a number of features of the GRBs that can be correlated with luminosity to calculate measured $d_L$ and distance modulus $m-M$ for comparison to cosmological predictions of distance modulus $m-M$.

The 69 data points of m-M distance modulus ($\mu^a$ vs Z of his Table 6) are shown in Fig. 15 with the theoretical SC-predicted curve. The fit is reasonable with the very large scatter. That scatter of the GB data is greatly amplified in the SC-$(H/H_0 - R/R_0)$ plot of the data of Fig. 16. That scatter at higher Z increases with increasing Z, even more rapidly than for the Riess SNIa data of Fig. 12.

Again assuming the astronomer can measure the flux of radiation more accurately than the redshift of its source, the SC-iterative HZ-process was also applied to these GRB-data and it likewise moved all of the data points to the theoretical curve and the $\Delta Z_i$ values were saved. However, when those $\Delta Z_i$ values were applied to the raw data of Fig. 15, all of the points were moved to the theoretical curve (HZF fit) except five which were driven to $Z \approx 0$ by very negative $\Delta Z$.

The HZF-process was then modified to operate on single data points with an added loop that subtracted $\Delta m = 0.1$ mag from the value of the distance modulus before the next calculation until the program reversed and did converge on reasonable $\Delta Z$. This modified program did work on all five points as shown in Fig. 17 where close inspection will show the five converged (HZF fit) points as small circles around a plus sign (instead of an X). The raw and the converged coordinates for these five points are listed in Table 1. No claim is made that these converged values are the correct values, but only that the converged values are consistent with, and support, the new theory.

**8.0 Other Features of New Cosmic Time**

Time stands out so prominent in Eq. (1). Figure 2 shows how cosmic time t varies with the expansion $R/R_0$. Eq. (A16) and Figs. 3 and 4 show the very tight inverse relation tH between time and the Hubble parameter. Are there other relations with this special SC-cosmic time that can be exhibited? There is another feature of cosmic time that can be extracted from Eq. (1) and from some of the earlier presentations.

The HZF process was successful in forcing a fit of SNIa and gamma ray data to agree with the putative correct theory. For this new feature, selected data from a number of other SNIa studies, including: Perlmutter, et al. [31]; Hamuy, et al. [32]; and Saha, et al. [33], are combined in Fig. 18 to show the theoretical relation of cosmic time between emission and measurement in our expanding universe. An apparent magnitude m $\approx 29$ represents at present the faintest object that can be detected [12]. Thus the various sets of HZ fit data points, with their reasonable distribution from HZF forced fit of a correct $H(Z)$, show support for theory over the entire range. Curve 1 comes very close to the left ordinate at m ~ 35 and that magnitude is not a linear scale, but a difference of five magnitudes corresponds to a factor of 100 in brightness. The Z of 1.7 occurs at t = 4.58 Gy and by t = 10 Gy $v_c$ is quickly approaching c around the bend to the near vertical drop to $v_c$ = c at t = $t_0$. So how is this curve of Fig 18 different than trajectory 1 of Fig. 7?

Trajectory 1 of Fig. 7 was for the real 4-D trajectory of radiation entering a telescope when it was pointing in one fixed direction. Remember that these data points in Fig. 18 were collected from the astronomer's telescopes pointing in many directions and the theoretical curve is the same on any planet in the universe of the same age $t_0$. The same supernova, if observed at the same $t_0$ from different planets, would appear at different positions on this same curve.

First consider curve 1. With increasing age $t_0$ of the universe, at least the lower part of the theoretical curve slowly moves from left to right and the "flashes" of radiation from different exploding stars appear and then disappear. To reason about the upper part of the curve, $(t \leq 10\,Gy)$ replace all of the points for SNIa with galaxies of equal but constant luminosity. All of the galaxies are moving away from the Earth with greater apparent magnitude m due to the expansion but, back with SNIa, $t_e$ doesn't change, so all of the theoretical curve (and points) must move up with increasing emission time t. So it appears that the $Z \cdot ED$ contribution of Fig. 9 raises the curve and it is the $c(t_0 - t_e)$ contribution part of Fig. 9 that moves forward the great bend of the curve.

To check the analysis, equations were added to transfer the base for our universe from the present size $R_0$ to a different size $R_0'$, $F = R_0'/R_0$, with new cosmological constants to calculate a new theoretical curve. Curve 2 at the earlier size of our universe, $F = 0.3621$ at $t_0' = 5.0$ Gy and curve 3 at the larger size $F = 1.606$ at $t_0' = 20.0$ Gy confirm the analysis and simple causality: $t_0' > t_e'$.

Given any supernova in the universe at redshift Z, the correct theory can predict the hypothetical measurement of m on any planet that the radiation happens to reach at $Z' = Z - \Delta Z$; or on a past or future size of our universe $R_0'$ that **the radiation happens to reach** at $Z'$ [6].

Theoretical curve 1 can be extended to ever higher m at lower emission time $t$. As astronomers build ever more powerful instruments, it will be interesting to see if any earlier $(2.0 \geq t_e', Gy \geq 0.2)$ higher-m sources continue to fit predicted curve 1.

**9.0 Summary and Conclusions**

This research began with the premise that present physics has completely missed the fundamental dynamic that brought our universe into existence and causes it to expand today. A few bold new physical concepts on space and time led to the concept that our universe is closed as an expanding 3-sphere. Much progress was made on the fundamental new concepts but not enough yet to constitute firm theoretical support for the beginning of our universe. The expansion after the beginning was complete so it was presented here as phenomenology without the fundamental missing dynamic.

Cosmic time appears in command of the expansion. In past physics, as a symmetric mathematical parameter, time appeared in a differential rate equation to be integrated. Now cosmic time appears explicitly squared as a product with one other variable, the total mass density of our universe; but one must know the ingredients of that mass density and how each scales with the expansion to drive the Universal Expansion, $G\rho t^2 = 3/32\pi$.

In SC-theory, *time* is given explicit dependence on $R$, $t(R)$, and is asymmetric as the positive root of Eq. (1) to match the unidirectional expansion with radial implicit dependence $R$ (t). For the content densities of our universe $\rho(R)$, present physics recognizes that of radiation and ordinary baryonic matter, and postulated dark matter that scales as $R^{-3}$. But this dark matter was replaced by a dark mass that scales with the expansion as $R^{-2}$, and gravitates but otherwise cannot interact directly with either radiation or matter.

Differentiation of expression Eq. (1) produced equations for the cosmological parameters of Hubble parameter H, expansion rate $\dot{R}/c$ and deceleration rate q. To complete the theory, the CMB temperature set the present radiation density $\rho_{r0}$ and nucleosynthesis calculations set the present matter density $\rho_{m0}$. WMAP measurements set the present age of the universe $t_0 = 13.5$ Gy to close the theory, because the present value of dark mass density is fixed by Eq. (1), $\rho_{x0} = \rho - \rho_{r0} - \rho_{m0}$. The theory then predicted present values of the cosmological constants: $H_0 = 68.6$ km s$^{-1}$ Mpc$^{-1}$, $q_0 = 0.0084$, and $\Omega_0(mass) = 0.28$, also in agreement with WMAP measurements.

Einstein's general relativity captured the local 3-D physics of clock-time in a 4-D geometry with block time as the fourth dimension. But an inadequate view of how our universe is constructed, as expressed in the Friedmann equations, was destined to lead to incorrect predictions of global phenomena of measurements of supernova Ia radiation that has traveled over long distances from the past.

A graphical sketch of a 3-sphere showed that physical 3-D velocities could be limited to the speed of light c while the radial velocity of expansion $\dot{R}$ could be normal to all 3-D velocities and could always be greater than c with future limit of $\dot{R} = c$. With the net velocity of radiation $v_c = Hr - c$ towards the astronomer's telescope, the graphic shows why current supernova Ia measurements are limited to $Z \leq 1.7$ because there $v_c = 0$ and beyond, the radiation is moving away but is rapidly being diluted in energy density by the expansion. Astronomers measure their received photon flux very accurately but all the very large expansion redshift effects must be accounted for to actually predict the measurements of the received flux radiation from these past distant objects.

The change of GR-dark matter, that scaled as $R^{-3}$, to SC-dark mass, that scales with the expansion as $R^{-2}$, not only allowed the derivation of Eq. (1) but also supports the postulate of a new substance that reacts only gravitationally with matter and reproduces as haloes around galaxies as $M_x(r) = M_{x0} r/r_0$ and thus flat rotational curves around spiral galaxies.

To properly account for the supernova Ia radiation, good measurements of the current cosmological parameters are not sufficient. One must know correctly the expansion rate $H(t)$ over the entire path of the radiation from emission until it enters the astronomer's telescope. Starting with no mechanism or physics of how our universe came to be, present GR-theory has been patched with early inflation, unknown dark matter, and later dark energy that accelerate the expansion. The measure of success of the GR-theory was how the apparent magnitude $m(Z)$ of the measured flux fit the predicted value using the measured spectral redshift z of the host galaxy. The deficiency of this test was shown by a new procedure where the theoretical $H(Z)$ for the data point in the $(\dot{a}/H_0) - a$ plane showed strong failure of all versions of the GR- $H(Z)$ theory.

In contrast, the SC-theory predicts a Hubble parameter $H(t)$ that correctly goes to zero in the future $(\dot{R}/R) \to c/R \to 0$ as $Z \to -1$ and predicts $H(Z)$ of the past to account for the supernova Ia data without dark energy and no acceleration of the universe.

The SC-theory predicted a curve of m-M or magnitude m(Z) about the same as the GR-curve. The significant difference came with the supernova Ia data plotted in the $H/H_0 - R/R_0$ plane. Here the data tended to follow the SC-theoretical curve and the data could be "HZF fit" to the theoretical curve in contrast to GR-theory where the data fit none of the GR-curves. Even the new gamma ray burst data out to Z = 6 could be "HZF" fit to SC-theory.

In conclusion, it has been argued that that the Hubble parameter $H$ as a function of cosmic time, t, implied by the SC-relation Eq. (1), is an accurate and useful representation of a variety of available data. In particular, H gives a good account of the SNIa data. It gives physically sensible predictions of what past and future astronomers could/would observe. It is hoped that this SC-relation will provide a signpost for future cosmological theory. The author will continue to develop and present deductions from the SC-theory.

## 10.0 ACKNOWLEDGEMENT

The author thanks his good friend, Emeritus Professor Robert A. Piccirelli, for extensive discussions of the new physical concepts.

TABLE 1
Change in Coordinates of Five GRBs

|  | BEFORE |  | AFTER |  |
| --- | --- | --- | --- | --- |
| **GRB** | **Z** | **m-M** | **Z** | **m-M** |
| 980613 | 1.10 | 45.85 | 1.47 | 45.43 |
| 060108 | 2.03 | 47.43 | 2.33 | 46.73 |
| 050406 | 2.44 | 47.92 | 2.72 | 47.17 |
| 050319 | 3.24 | 48.06 | 3.50 | 47.88 |
| 071214 | 3.42 | 48.44 | 3.66 | 48.01 |

# APPENDIX
## Summary of the SC-Cosmological Theory

The scale factor R has units of length for our 3-sphere, spatially 3-dimensional expanding universe; G is the gravitational constant; c is the local speed of light; and H is the Hubble parameter. Present values have subscript 0 and cgs units are assumed. Other subscripts include: r=radiation, m=matter and x=dark mass (not dark matter). Pertinent equations of the new theory [hereafter: "SC-Theory"] are listed in Table A.

Table A Derivation

| | | |
|---|---|---|
| Universal constant | $\kappa = Gt^2 \rho = Gt_0^2 \rho_0 = 3/32\pi$ | (A1) |
| From $T_0=2.726$ K | $\rho_{r0} = 9.40 \times 10^{-34}$. | (A2) |
| From nucleosynthesis: | $\rho_{m0} = 2.72 \times 10^{-31}$ | (A3) |
| Present age, (Input): | $t_0 = 13.5$ Gy | (A4) |
| From (A1): | $\rho_0 = (\kappa/G)/t_0^2$ | (A5) |
| From (A6): | $\rho_{x0} = \rho_0 - \rho_{r0} - \rho_{m0}$ | (A6) |
| From above | $R_0 = ct_0 (\rho_0/\rho_{x0})^{1/2}$ | (A7) |
| Redshift Z (Input): | $(1+Z) \equiv R_0/R$ | (A8) |
| Radiation at Z: | $\rho_r = \rho_{r0}(R_0/R)^4 = \rho_{r0}(1+Z)^4$ | (A9) |
| Matter at Z: | $\rho_m = \rho_{m0}(R_0/R)^3 = \rho_{m0}(1+Z)^3$ | (A10) |
| Dark Mass at Z: | $\rho_x = \rho_{x0}(R_0/R)^2 = \rho_{x0}(1+Z)^2$ | (A11) |
| Total at Z: | $\rho(R) = \rho_r + \rho_m + \rho_x$ | (A12) |
| Cosmic time: | $t(R) = +(t_0^2 \rho_0/\rho(R))^{1/2}$ | (A13) |
| From derivatives, d/dt: | $\rho_2 = 2\rho_r + (3/2)\rho_m + \rho_x$ | (A14) |
| From derivatives, d/dt: | $\rho_3 = 4\rho_r + (9/4)\rho_m + \rho_x$ | (A15) |
| From derivatives, d/dt: | $H = \dot{R}/R = (\rho/\rho_2)/t$ | (A16) |
| Expansion Rate: | $\dot{R}/c = (R/ct)(\rho/\rho_2)$ | (A17) |
| Deceleration, $q = -\ddot{R}R/\dot{R}^2 =$ | $\left(-1 + \left(3 - 2(\rho\rho_3/\rho_2^2)\right)/Ht\right)$ | (A18) |

The scaling with the expansion of radiation, Eq. (A2), and matter, Eq. (A3), are borrowed from the big bang model, as is the value of $\kappa$ for early Friedmann radiation.

The postulated scaling, Eq. (A11), of the new and now dominant stuff called "dark mass," is the key signature of this new cosmological model. Its density decreases with the expansion but its total mass, always in individual clumps, increases with the expansion. It is not a 3-D substance and so does not interact with radiation or matter except gravitationally, where it contributes to the local curvature of 3-D space. The distribution of these miniscule dark mass seeds at the beginning of the expansion sets the pattern for the present large-scale structure, including voids, and contributes to the early formation of black holes and fit to supernova Ia data for $t_0=13.5$ Gy with no acceleration of the expansion rate.

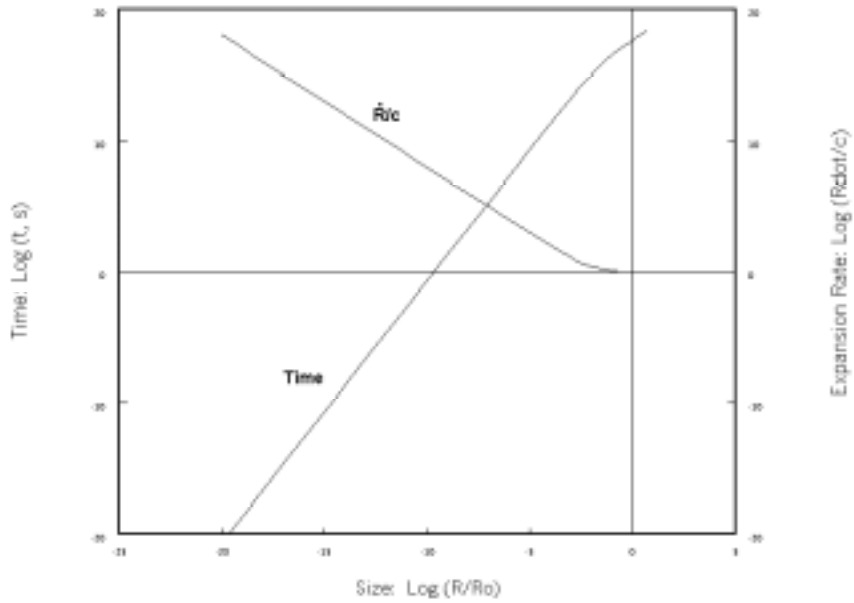

Fig. 1 Time has passed through radiation and matter domination and into the final dark mass domination. The dimensionless expansion rate $\dot{R}/c$ of our universe has now almost reached its limiting value of unity or R = ct

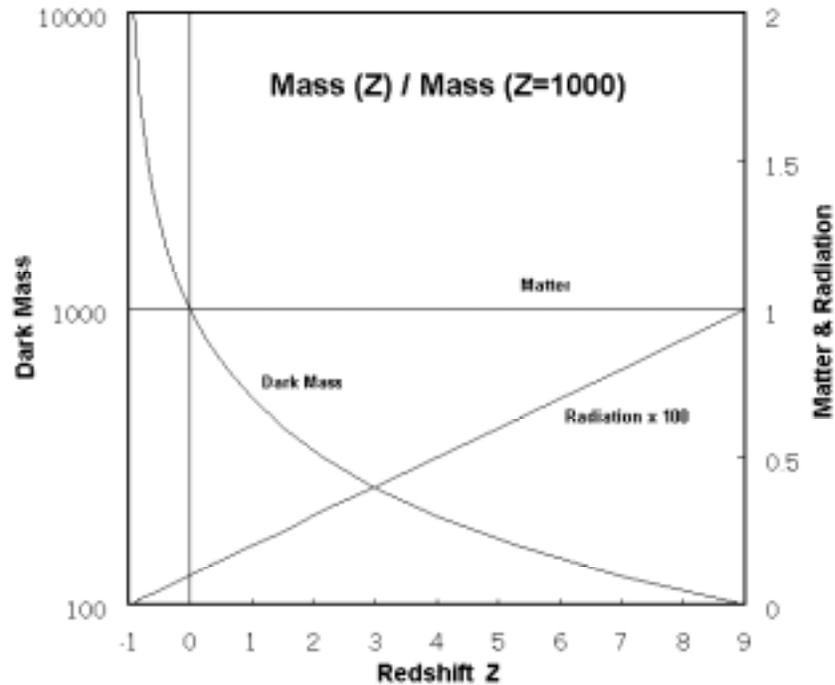

Fig. 2 The change in the three components of the total mass of the universe is shown by their respective ratios to their component mass at decoupling at Z ≈ 1000. The future rapid rise of SC-dark mass will increase condensation of matter into galaxies and clusters.

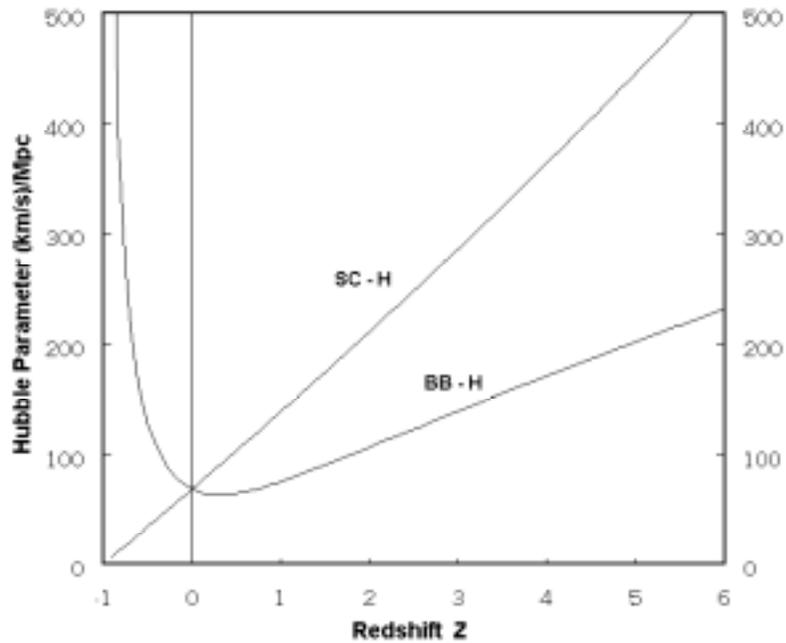

Fig. 3 In the SC-theory all densities go to zero in the far future as does the Hubble parameter SC-H. The big bang H was also headed toward zero but the added energy of the cosmological constant to make $\Omega = 1$ creates the acceleration with an unphysical drive of the BB-H curve to infinity in the far future.

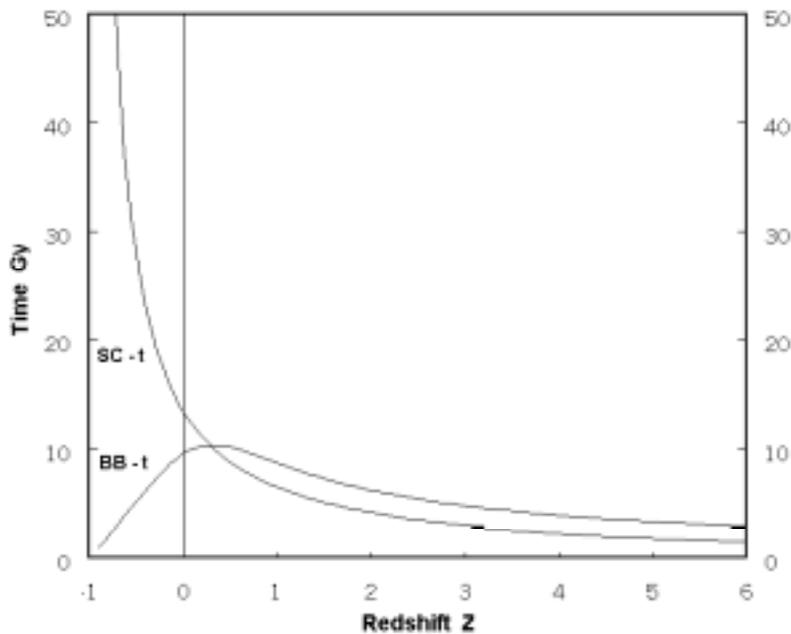

Fig. 4 Both SC-theory and BB-theory predict $tH \to 1$ in the far future. Thus time is driven in the opposite direction to that of H in Fig. 5. Here SC-time increases without limit as the universe expands as we expect, but the added acceleration to the BB-model reverses BB-time with an unphysical drive to zero in the far future.

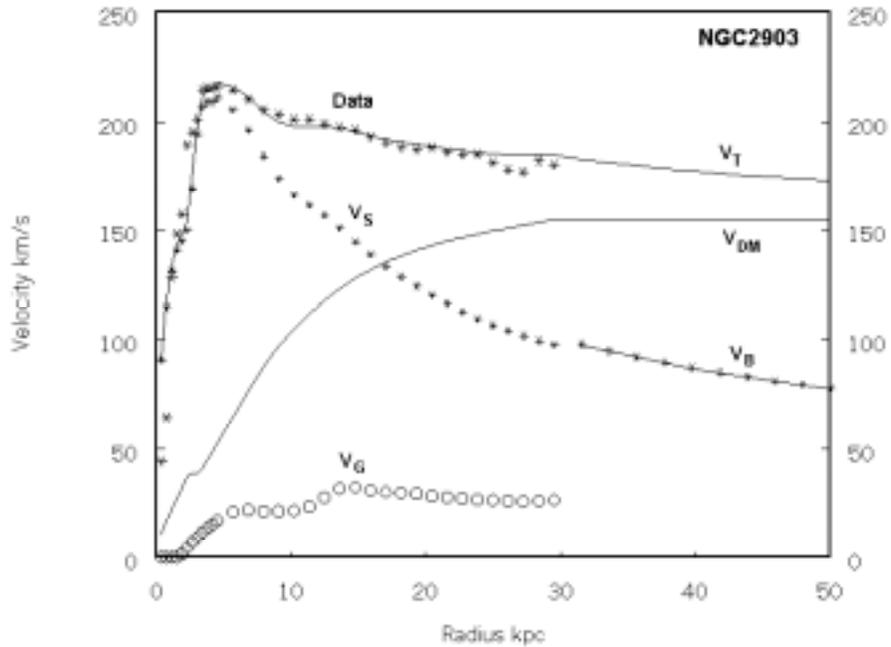

Fig. 5 The MOND-acceleration $a_0$, plus dark mass scaling $r^{-2}$, is combined (see text) to predict the rotation curve with predominant matter in the core ($V_S$ = stars and $V_G$ = gas). Rapid rise time of rotation curve indicates an early galaxy ($V_{DM}$ =dark mass).

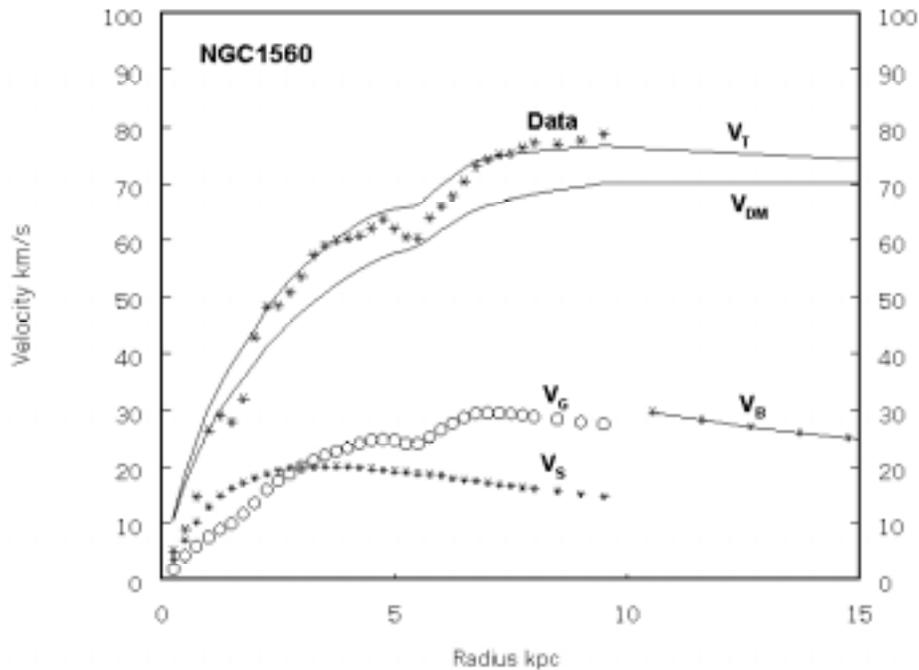

Fig. 6 The MOND-acceleration $a_0$, plus dark mass scaling $r^{-2}$, is combined (see text) to predict the rotation curve with little matter, but predominant dark mass, in the core ($V_{DM}$ = dark mass).

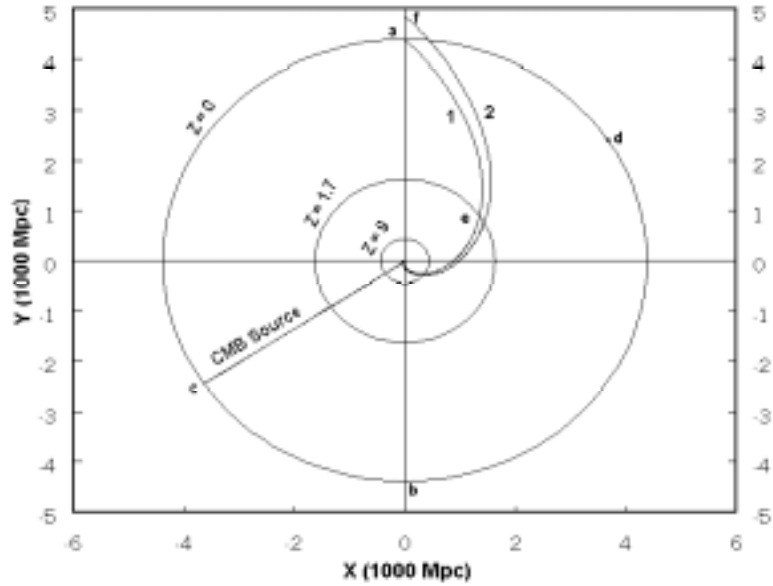

Fig. 7 Radiation (v = c) must travel on the outer surface (3 circles) of the expanding 4-D ball. The 4-D radius, expanding faster $\dot{R} > c$ (but decelerating) produces the spiral trajectory from emission to detection in the telescope at point a. The stationary 3-D CMB source starts at Z = 999 and moves on 4-D $R$ from near the origin o to point c.

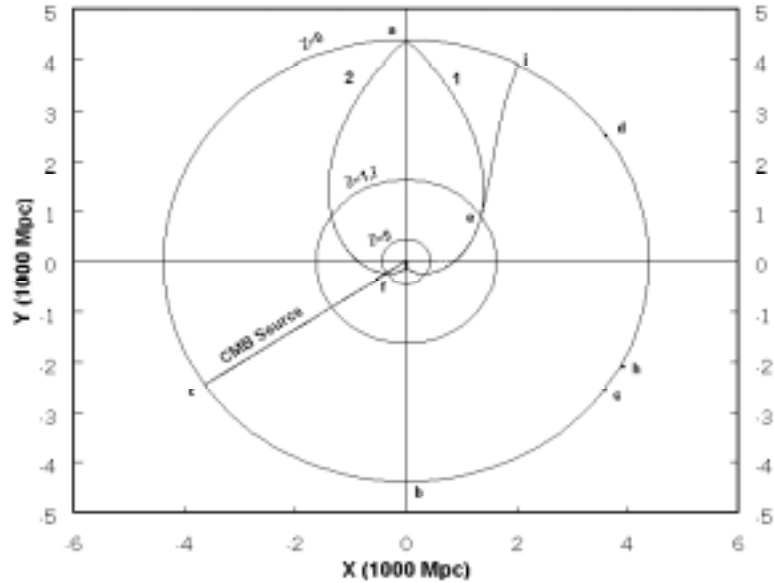

Fig. 8 Pointing the telescope in the opposite direction receives the radiation from trajectory 2. Now and into the future, redshift Z = 1.7 has the amazing distinction of always being that past size of the universe $R = R_0/2.7$ where the net velocity $v_c = HR - c = 0$ toward (past world line of) the telescope. The trajectory through point "i", show s the result of added acceleration of constant H at Z = 1.8.

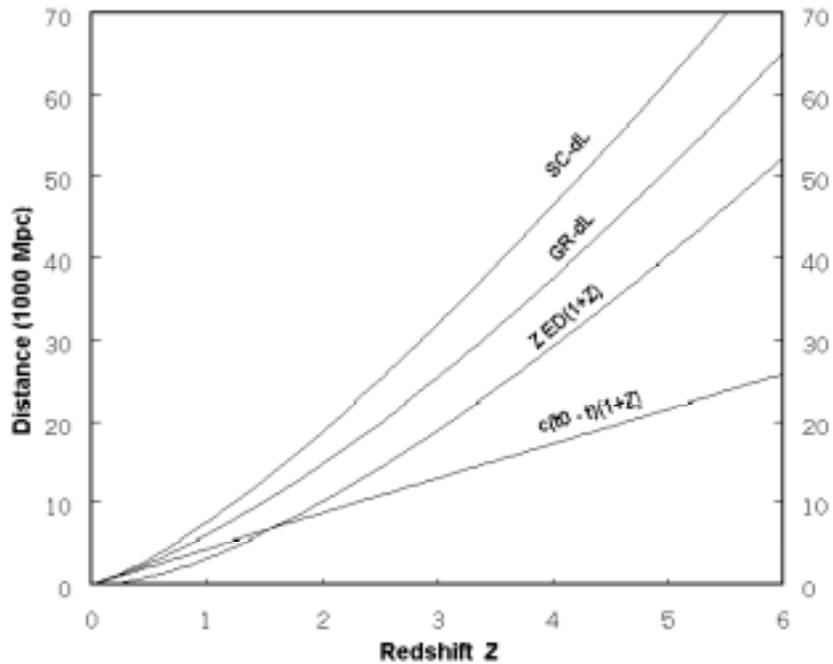

Fig. 9 SC-$d_L$ predicts greater luminosity distances than GR-$d_L$ and therefore greater apparent magnitude m in the range $1 \leq Z \leq 2$ and so no need for GR-dark energy and acceleration of the expansion. The SC-$d_L$ is the SUM of velocity of light $c(t_0 - t_e)$ and expansion $Z \cdot ED$ components.

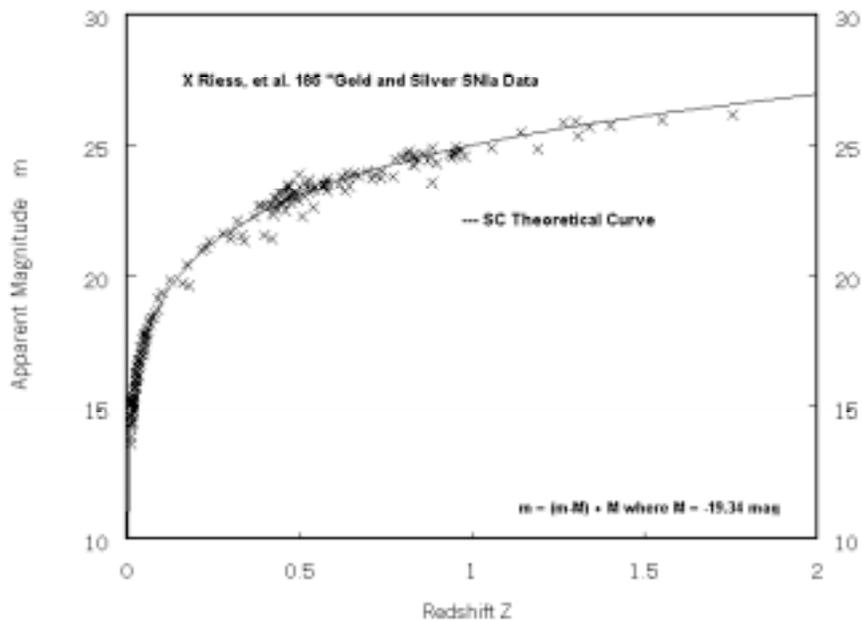

Fig. 10 Riess et al, 185 "gold and silver" (m-M) + [M = -19.34] are compared to the SC-theoretical curve for m. The scatter is too great to select between competing theories in this type of plot.

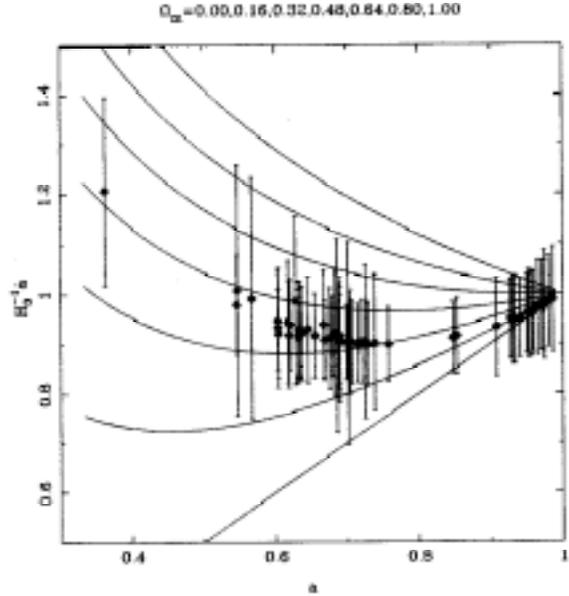

**Figure 3.** The observed supernova data points in the $\dot{a} - a$ plane for flat models. The procedure for obtaining the data points and the corresponding error-bars are described in the text. The solid curves, from bottom to top, are for flat cosmological models with $\Omega_m = 0.00, 0.16, 0.32, 0.48, 0.64, 0.80, 1.00$ respectively.

This Fig. 3 of astro-ph/0212573 was reproduced with the kind permission of author T. R. Choudhury.

Fig. 11 The two important points are: (1) the data tend not to follow any one of the seven big-bang theoretical curves for $\Omega_m$; and (2) this new procedure of plotting in the $\dot{a}$ - a plane is most useful in analyzing the SC-equations in the equivalent $H/H_0 – R/R_0$ plane.

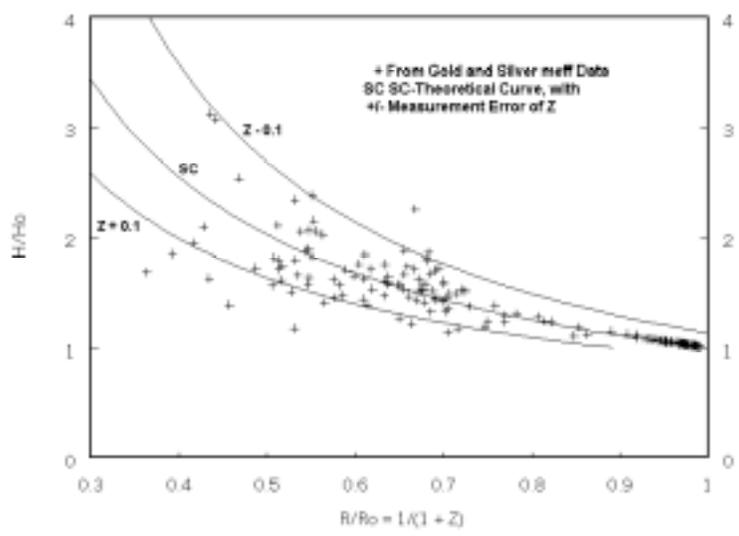

Fig. 12 With reported data m(z) converted to values of $H(m, Z)$, the SC-theory correctly predicts the trend of the data and even the increasing scatter of the data with increasing redshift Z. At $R/R_0 \approx 0.3$, SC-$H/H_0$ is about a factor of three greater than in Fig. 11.

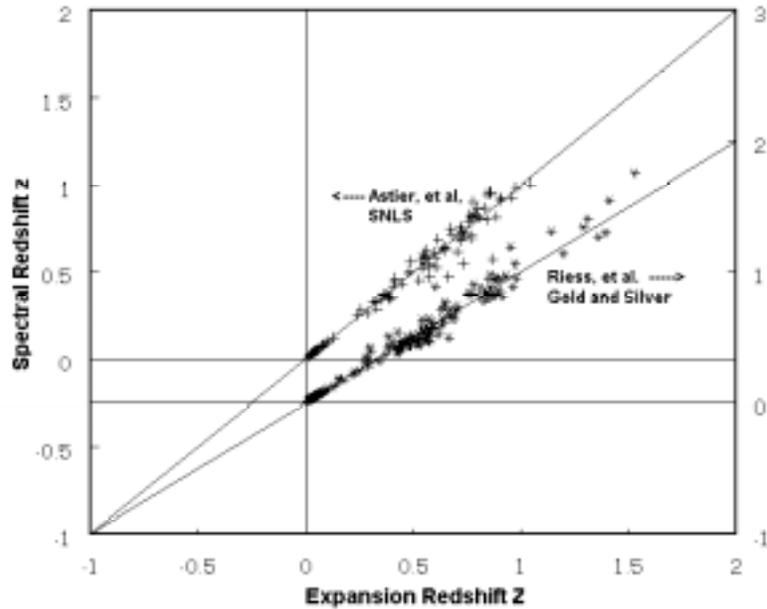

Fig. 13 Good theory allows measured spectral redshift z of the host galaxy to agree with expansion redshift $Z = R_0/R - 1$ determined by the measured flux F of radiation from the source. Here theory dependent reported data can contaminate such a plot (see text)..

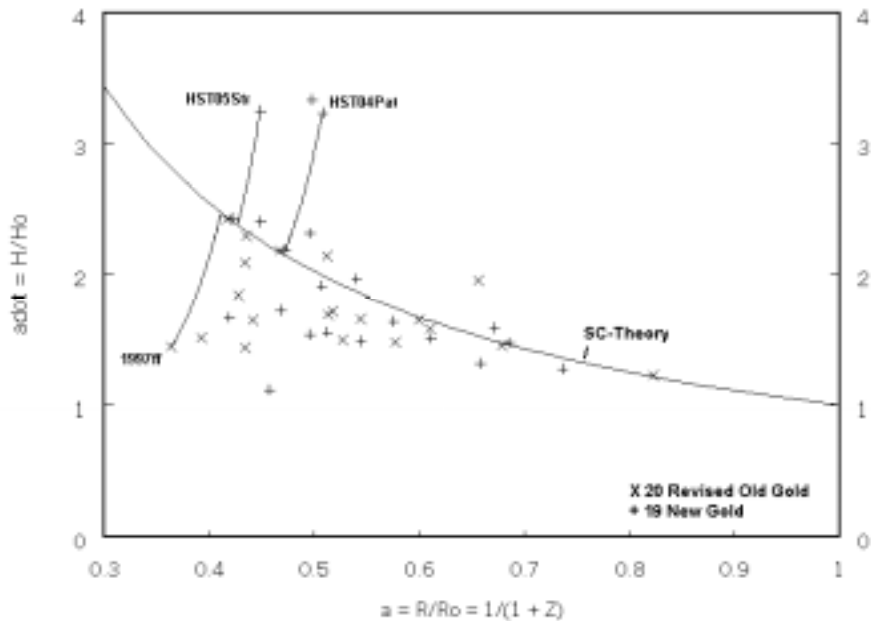

Fig. 14 Twenty of the SNIa of Fig. 12, with upgraded distant modulus m-M, have here poorer average deviation from SC-H(Z) (see text). The new 19 SNIa did better. Calculations were followed to demonstrate the force-fit to H(Z) of three high-Z SNIa. Increasing Z, lowers a and decreases H, decreasing Z, increases a and increases H.

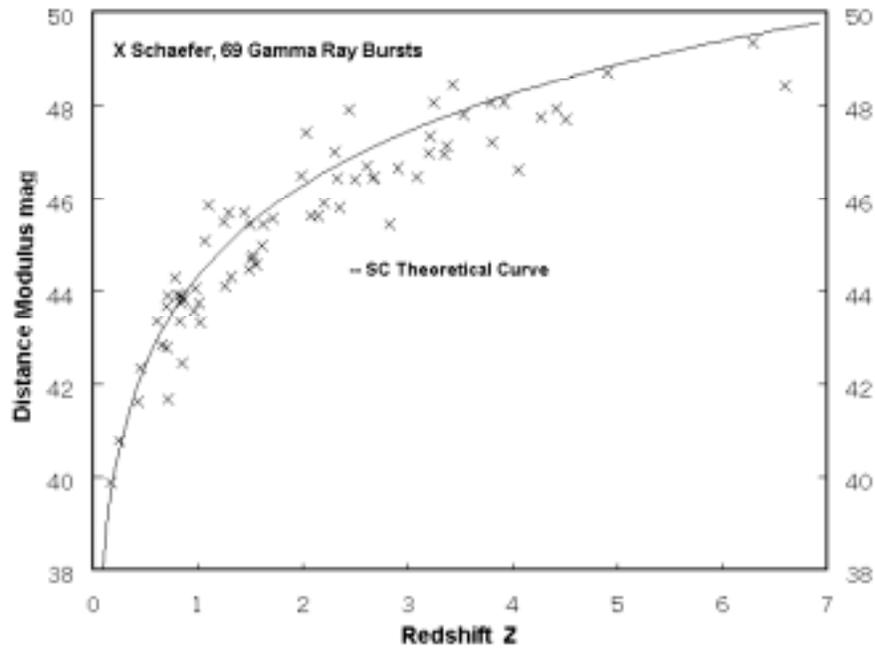

Fig. 15 Gamma ray bursts can be seen to much higher redshift Z than supernova Ia. Even with greater scatter, features of their spectra indicate they might provide "standard candles" for exploring R(t) or H(Z) deeper into our past universe,

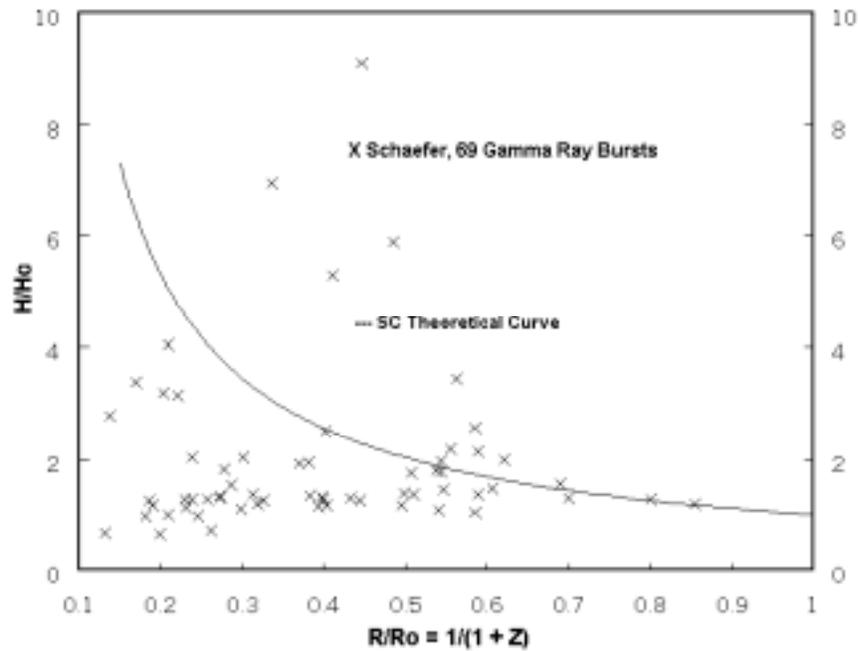

Fig 16 As with supernova Ia data, the scatter of the data is greatly increased (even more at greater Z) in the $H/H_0$ – $R/R_0$ plane. Nevertheless, the SC-theoretical curve seems to guide the general increase in H(Z).

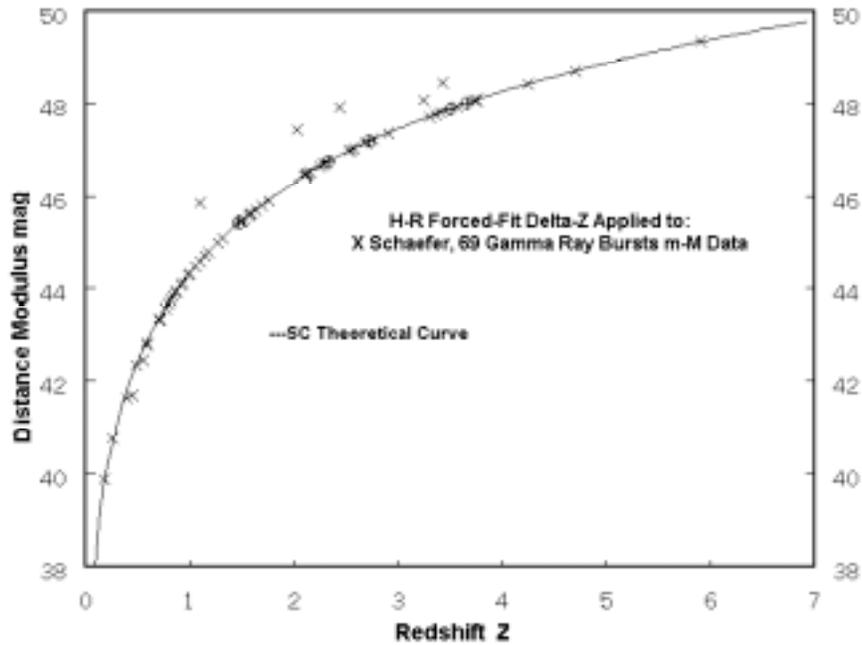

Fig. 17 The $(\Delta Z)_i$ to force the fit in Fig. 17 were applied to the data in Fig 15 and again all of the data points were driven to fit the SC-theoretical curve except for the five shown (driven toward Z = 0). After a small decrease in m-M (see text), these five were also driven to the points on the curve with labels of "plus inside a circle".

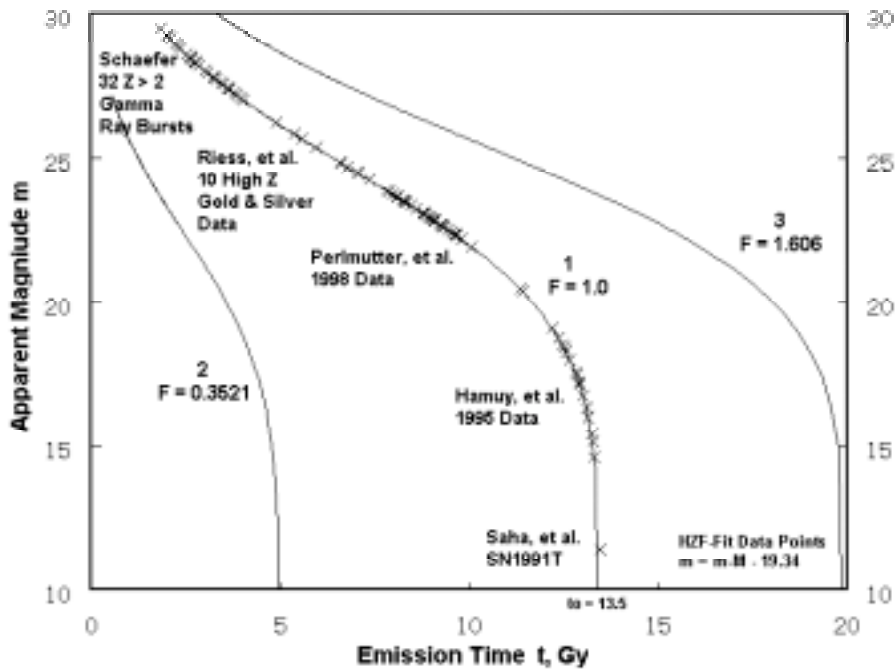

Fig. 18 In its straight-line path from source to detector, the energy density of radiation is degraded by many physical effects including $1/r^2$ loss, relativistic and expansion effects. Here the SC-theory of predicted SNIa magnitude versus time of emission is shown with superimposed "HZF Fitted" data from a number of studies (see text).